\documentclass[11pt]{article}
\pdfoutput=1
\usepackage{jheppub}
\usepackage{graphicx}
\usepackage{amssymb}
\usepackage{bbm}
\usepackage{microtype}
\usepackage{mathrsfs}
\usepackage{amsmath,amssymb}
\usepackage{slashed}
\usepackage{hyperref}
\usepackage{circuitikz}
\usepackage{caption}
\usepackage{float}
\usepackage{xcolor}
\usepackage{dsfont}
\usepackage{verbatim}
\usepackage{subfig}
\usepackage{amsmath,amssymb,amscd,amsfonts,mathtools,csquotes}
\usepackage{xifthen}
\usepackage{xcolor}
\definecolor{markgreen}{RGB}{230,243,230}
\definecolor{darkolivegreen}{rgb}{0.33, 0.42, 0.18}
\definecolor{darkpastelgreen}{rgb}{0.01, 0.75, 0.24}

\usepackage[toc,page]{appendix}
\usepackage{epsfig}
\usepackage{epstopdf}
\usepackage{latexsym}
\usepackage{graphicx}
\usepackage{booktabs}
\usepackage{bbm}
\usepackage{color}
\usepackage{physics}
\usepackage{tensor}
\usepackage{verbatim}
\usepackage{subfig}
\usepackage{tikz}
\usepackage{ifthen}
\usepackage{array}
\usetikzlibrary{matrix}
\usetikzlibrary{decorations.markings,calc,shapes,decorations.pathmorphing,patterns,decorations.pathreplacing,arrows,arrows.meta}
\usetikzlibrary{positioning}
\usepackage{hyperref}
\usetikzlibrary{patterns}
\usetikzlibrary{positioning}
\newdimen\mydim
\pdfoutput=1
\makeatletter
\def\@fpheader{\relax}

\newcommand*{\ov}[1]{
  $\m@th\overline{\mbox{#1}}$
}
\newcommand*{\ovA}[1]{
  $\m@th\overline{\mbox{#1}\raisebox{3mm}{}}$
}
\newcommand*{\ovB}[1]{
  $\m@th\overline{\mbox{#1\rule{0pt}{3mm}}}$
}
\newcommand*{\ovC}[1]{
  $\m@th\overline{\mbox{#1\strut}}$
}
\newcommand*{\ovD}[1]{
  $\m@th\overline{\mbox{#1\vphantom{\"A}}}$
}
\newcommand*{\ovE}[1]{
  $\m@th\overline{\raisebox{0pt}[1.2\height]{#1}}$
}
\newcommand*{\ovF}[1]{
  $\m@th\overline{\raisebox{0pt}[\dimexpr\height+1mm\relax]{#1}}$
   Package `calc' can be used as alternative for `\dimexpr'.
}
\newcommand*{\ovG}[1]{
  $\m@th\overline{\raisebox{0pt}[\dimexpr\height+1mm\relax]{#1\vphantom{A}}}$
}
\makeatother
\newboolean{showcomments}
\setboolean{showcomments}{false}
\newcommand\rem[1]{\ifthenelse{\boolean{showcomments}}{{#1}}{}}
\newcommand{\be}{\begin{equation}}
\newcommand{\ee}{\end{equation}}

\usepackage{xifthen}
\newcommand{\dalembert}[1][]{\ifthenelse{\isempty{#1}}{\Box}{#1\Box}}
\usetikzlibrary{decorations.pathmorphing}
\tikzset{snake it/.style={decorate, decoration=snake}}

\title{\Large Closed Universe, Quantum Entanglement and Observers}
\author{Hao Geng}
\affiliation{Gravity, Spacetime, and Particle Physics (GRASP) Initiative, Harvard University, 17 Oxford St., Cambridge, MA, 02138, USA.}
\emailAdd{haogeng@fas.harvard.edu}

\abstract{It has been argued that in quantum gravity the Hilbert space associated with any closed universe is trivially one-dimensional. To reconcile this argument with the richness of our own universe, it was suggested that a nontrivial Hilbert space will emerge once an ``observer" is included. Such an ``observer" has been modeled in different ways. There are three concrete classes of models: \textbf{1)} The ``observer" is modeled by a term in the full Hamiltonian of the system which is linear in a phase space variable; \textbf{2)} The ``observer" is modeled in the quantum circuit model of holography as a subsystem in the closed universe which is entangled with an external non-gravitational system; \textbf{3)} The ``observer" is argued to be described by the emergent Goldstone bosons when the diffeomorphisms are spontaneously broken. In fact, the third class was discovered by an attempt to physically realize the first class. In this paper, we will show that the second class can also be unified into the third class. Our result indicates that the emergence of the ``observer" is a manifestation of the \textit{Massive Islands} paradigm. It demonstrates that the graviton mass is a feature of the quantum state. One should think of the islands as closed universes and the Goldstone boson associated with the graviton mass as the ``observer", which is exactly the case at late-times for evaporating black holes. An implication of this work is that the ``observer" can ``observe" global symmetries in quantum gravity.}

\begin{document}
\maketitle
\flushbottom
\pagebreak

\section{Introduction}
The modern program of quantum gravity started with the attempt to understand the physics of black holes and their Hawking radiation \cite{Hawking:1974sw}. There are two reasons for such efforts. First, black holes seem to be very puzzling if one thinks about various quantum aspects of them from a naive low-energy effective theory perspective \cite{Hawking:1976ra,Mathur:2008nj,Mathur:2009hf,Almheiri:2012rt,Almheiri:2013hfa}. Second, the interior of the black hole is in fact an Friedman-Robertson-Walker (FRW) universe with (almost) closed spatial 
slices and this is similar to the situation of our own universe. Thus, understanding black holes would help us see how gravity works quantum mechanically, and it is potentially useful for making predictions about our own universe in the very quantum regime.

The discovery of the AdS/CFT correspondence \cite{Maldacena:1997re,Gubser:1998bc,Witten:1998qj} enabled the studies of quantum aspects of black holes within low-energy effective field theories. During these studies it has been realized that intrinsic gravitational effects play an essential role in resolving various puzzles, such as \cite{Hawking:1976ra,Mathur:2008nj,Mathur:2009hf,Almheiri:2012rt,Almheiri:2013hfa}, and interestingly all these effects are captured by low-energy effective field theories \cite{Papadodimas:2012aq,Papadodimas:2013jku,Raju:2018zpn,Almheiri:2019psf,Almheiri:2019qdq,Penington:2019npb,Penington:2019kki,Geng:2020qvw,Raju:2020smc,Geng:2021hlu,Raju:2021lwh,Geng:2023zhq,Geng:2024xpj,Geng:2025rov}. This is reasonable, as all of those puzzles are based on our intuition in low-energy effective field theory, where there is a fixed spacetime background with gravity weakly coupled. Therefore, the resolutions of the puzzles should live in the same regime.\footnote{With all physical effects properly incorporated.} 

However, attempting generalizations of the above lessons we learned about black holes, together with some extra seemingly plausible inputs from string theory, suggest that the Hilbert space associated with any closed gravitational universe must be trivially one-dimensional \cite{McNamara:2019rup,McNamara:2020uza,Harlow:2025pvj}. This conclusion in principle could apply to our own universe and it is thus in apparent conflict with our everyday experience of the richness in our own universe. Hence, various works have been published questioning if the assumptions leading to the above conclusion make sense \cite{Chakraborty:2023los,Chakraborty:2023yed,Chakraborty:2025izq} or if there is any way to relax the above conclusion \cite{Chandrasekaran:2022cip,Witten:2023qsv,Harlow:2025pvj,Balasubramanian:2023xyd,Maldacena:2024spf,Abdalla:2025gzn,Nomura:2026igt}.\footnote{See also \cite{Klinger:2023auu,Klinger:2023tgi,Klinger:2023auu,AliAhmad:2024saq,AliAhmad:2024eun,AliAhmad:2024wja,Klinger:2025hjp,Cirafici:2026iek,Das:2026npy,Cheung:2026euf,Seo:2026kxs,Nomura:2026igt,Kaushal:2025wbn,Chakraborty:2025baw,Ciambelli:2025ztm,Chakraborty:2025izq,Aalsma:2025bcg,Guijosa:2025mwh,AliAhmad:2025ukh,Jensen:2024dnl,Piazza:2024lpt,Cirafici:2024itu,Grassi:2024vkb,Blommaert:2025bgd,Liu:2025cml,Liu:2025ikq,Liu:2026fnd,McNamara:2026isz,Blommaert:2026lvp,Blommaert:2026ofx,dai2026babyuniverselargefamily,giddings2026leadinggravitationaldressingoperators} for relevant works.} An emerging consensus in the literature is that if the above conclusion is correct, it could be relaxed if there is some ``observer" that lives inside the closed universe and the Hilbert space observed by this ``observer" could be nontrivial. 

Such an ``observer" was expected to model ourselves and was given a fundamental role in the theoretical predictions of the results of the observations we would make. This sounds rather philosophical, if not religious. A plausible assertion that attempts to make sense of this idea is the following: 
\begin{displayquote}
\textit{In a closed universe, any observer is necessarily part of the system. Hence, a full description of this system, such as the theory of quantum gravity, should necessarily be conditioned on the existence of observers. So, it is not surprising that the physics inside this closed universe, i.e., the physics that could be observed by an observer, could be very different from the physics without any observer.}    
\end{displayquote}
However, such an ``observer" has been modeled in very different ways to generate nontrivial physics \cite{Chandrasekaran:2022cip,Witten:2023qsv,Witten:2023xze,Harlow:2025pvj}, and it was an open question how those models could be physically realized. This problem has been pointed out and studied in \cite{Geng:2024dbl,Geng:2025bcb,Geng:2025gqu}, where it was realized that the ``observer" as proposed in \cite{Chandrasekaran:2022cip,Witten:2023qsv,Witten:2023xze} can be thought of an oversimplified model of a more general physical phenomenon. This phenomenon is the Higgs mechanism, and the conclusion that the Hilbert space is trivial in a closed gravitational universe is analogous to the fact that in quantum electrodynamics there cannot be an isolated charged particle on a closed manifold unless the gauge symmetry is spontaneously broken.\footnote{We note that gauge symmetry is not physical and it is never broken. What we really mean is the global part of the gauge symmetry is spontaneously broken. In the discussion of this paper, we will ignore this subtlety and simply say that the gauge symmetry is spontaneously broken for the sake of convenience.} More precisely, it was realized that when the diffeomorphisms are spontaneously broken, the emergent Goldstone vector boson will play the role as the ``observer" in the sense of \cite{Chandrasekaran:2022cip,Witten:2023qsv,Witten:2023xze}. Intuitively, this is the quantum version of the classical observation that in our universe there are many nontrivial features, such as a non-uniform distribution of galaxies or an inflationary period back in the days, which spontaneously break the diffeomorphisms. These features can be used as rods and clock to define a coordinate system, and one can describe other observables in this coordinate system, i.e. relative to these features. Interestingly, it was noticed in \cite{Geng:2025gqu,Geng:2025gns,Geng:2025byh} that, as opposed to the above intuitive example with semiclassical features, such features could also be intrinsically quantum, i.e. features of the quantum state such as entanglement. As will be discussed in Sec.~\ref{sec:HUZ}, the emergent ``observer" from the quantum features is in fact a physical realization of the type of ``observer" proposed in \cite{Harlow:2025pvj}. Hence, the above result that the ``observer" emerges as the Goldstone boson from the Higgs mechanism associated with the spontaneous diffeomorphism breaking provides a unified underlying description of the existing proposals of the ``observer" in the literature. We will provide an explicit model of such a closed gravitational universe in Sec.~\ref{sec:Construction} and show that indeed there is an ``observer", and, as a consequence, the Hilbert space in it is necessarily nontrivial.

\begin{figure}[H]
\begin{centering}
\begin{tikzpicture}[scale=0.6,decoration=snake]
       \draw[-,very thick] 
       decorate[decoration={zigzag,pre=lineto,pre
       length=5pt,post=lineto,post length=5pt}] {(-2.5,0) to (2.5,0)};
       \draw[-,very thick,black] (-2.5,0) to (-2.5,-7);
       \draw[-,very thick,blue] (5,3) to (5,-7);
         \draw[-,very thick] 
       (-2.5,-7) to (5,-7);
       \draw[-,very thick,red] (2.5,0) to (-2.5,-5);
        \draw[-,very thick,black] (2.5,0) to (2.5,3);
        \draw[-,very thick,green] (5,3) to (10,-2);
         \draw[-,very thick,green] (10,-2) to (5,-7);
       \draw[fill=green, draw=none, fill opacity = 0.1] (5,3) to (10,-2) to (5,-7) to (5,3);
       \draw[-,very thick,black] (-2.5,0) arc (90:0:1.5 and 7);
\draw[thick,black,pattern=north east lines,pattern color=black] (-2.5,0) arc (90:0:1.5 and 7)--(-2.5,-7)--(-2.5,0);
\draw[-,very thick,black!20!orange] (10,-2) ..controls (5,1.1).. (2.5,1);
\draw[-,very thick,black!20!orange] (-2.5,-1.5)..controls (0,-1.5)..(2.5,0);
\end{tikzpicture}
\caption{\small A demonstration of the late-time Cauchy slice of an evaporating black hole in AdS. The black hole forms from some matter (the shaded grey region) collapse. The shaded green region is a non-gravitational bath that induces the black hole evaporation. The union of the orange slices is a late-time Cauchy slice with the part in the black hole interior almost a closed manifold.}\label{pic:interior}
\end{centering}
\end{figure}
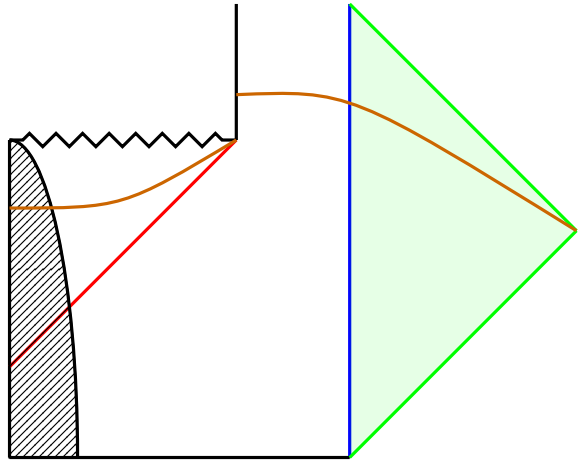

We would like to emphasize that, as discussed at the beginning of this introduction section, the above understanding of the physics of closed gravitational universes is in fact a generalization of an important lesson that we have learned in the study of black holes. This is regarding the \textit{entanglement island}, which by definition is a subregion in a gravitational universe whose physics is fully encoded in an external non-gravitational system away from it. Entanglement islands emerge in the study of the black hole information paradox, where one is attempting to compute the Page curve of the emitted black hole radiation. In that context, the entanglement islands can be thought of as the black hole interior after the Page time, with that external non-gravitational system as the early-time Hawking radiation emitted by the black hole. As a result, the physics inside the black hole is fully captured by the early-time Hawking radiation. This gives the unitary Page curve and provides a resolution of the information paradox \cite{Almheiri:2019psf,Penington:2019npb,Almheiri:2019yqk,Laddha:2020kvp,Geng:2020fxl,Raju:2020smc,Geng:2021hlu}. In fact, one should think of the entanglement islands as prototypes of closed gravitational universes. This connection manifests itself in one example---evaporating black holes at late-times. In this case, the black hole interior would be the late-time entanglement islands of all the emitted Hawking radiation, and in fact the geometry of the black hole interior is almost a closed universe, as the constant-time slices\footnote{Note that in the Schwarzschild coordinates, the interior time is the exterior radial coordinate $r$.} are almost closed manifolds (see Fig.~\ref{pic:interior}).\footnote{A potential subtlety is the black hole singularity which unfortunately is the edge of the interior spatial slices. Since we are interested in problems in the low energy effective theory, we assume that the drama about the singularity is not important in this regime. Modulo this subtlety, one can take the interior spatial slices as closed.} Even though in general cases there are spatial boundaries of entanglement islands, this connection is still fine because in a fully gravitating spacetime, where gravity is everywhere dynamical, the boundary of a finite subregion is not so different from the interior. This is because there are no extra modes that purely live on the boundary. The decreasing part of the Page curve suggests that the Hilbert space associated with an entanglement island is necessarily nontrivial, as it indicates nontrivial information transfer between the black hole interior and exterior. In \cite{Geng:2020qvw,Geng:2020fxl,Geng:2021hlu,Geng:2023zhq} it has been pointed out that the setups which contain entanglement islands in the above attempt to calculate the Page curve of the Hawking radiation are in fact in the context of massive gravity. Later, it was understood in \cite{Geng:2025rov,Geng:2025byh,Geng:2025gqu,Geng:2026asi} that the graviton mass is essential for entanglement islands and it originates from the Higgs mechanism associated with the diffeomorphism. Furthermore, the feature that spontaneously breaks the diffeomorphisms is the quantum entanglement between the black hole spacetime and the external non-gravitational system that was used to collect the black hole radiation. The associated Goldstone boson ensures the consistency of observables inside the entanglement islands, and it is exactly the emergent``observer" as we discussed above in the context of closed universes \cite{Geng:2025bcb}. The existence of the nontrivial observables, ensured by the Goldstone boson (``observer"), indicates the nontriviality of the Hilbert spaces associated with entanglement islands (closed universes).

The remaining parts of this paper are organized as follows. In Sec.~\ref{sec:Construction}, we revisit the physics of the Karch-Randall braneworld and provide an explicit example of a closed gravitational universe in the Karch-Randall braneworld. In Sec.~\ref{sec:observer}, we analyze the gravitational physics in the resulting closed universe, show the emergence of the ``observer" and explain why it guarantees that the Hilbert space associated with the closed universe is necessarily nontrivial. We will also explain why the origin of the ``observer" is quantum entanglement. In Sec.~\ref{sec:comments}, we will comment on the connection between our study and the existing literature, discuss implications of our current work and refine various lessons we learned about quantum gravity in the study of black holes. We will conclude this paper with further discussions in Sec.~\ref{sec:conclusion}.

\section{Constructing the Closed Universe in the Karch-Randall Braneworld}\label{sec:Construction}
The Karch-Randall braneworld \cite{Karch:2000ct,Karch:2000gx,Geng:2026wok} has been a useful setup to generate solvable models for deep questions in quantum gravity, e.g., the Page curve of black hole radiation \cite{Almheiri:2019hni,Geng:2020qvw,Chen:2020uac,Chen:2020hmv}. Meanwhile, it provides additional insights into the nature of quantum gravity, such as how information is localized and encoded in a gravitational universe \cite{Almheiri:2019psy,Almheiri:2019hni,Geng:2020qvw,Geng:2020fxl,Chen:2020uac,Chen:2020hmv,Geng:2021hlu,Geng:2023qwm,Geng:2025rov,Callebaut:2026hso} and whether there could be global symmetries \cite{Anastasi:2025puv,Geng:2025gns}.

In the simplest cases, the Karch-Randall braneworld is described by the following action
\begin{equation}
S=-\frac{1}{16\pi G_{N}}\int_{\mathcal{M}_{d+1}} d^{d+1}x\sqrt{-g} (R-2\Lambda)-\frac{1}{8\pi G_{N}}\int_{\mathcal{\partial M}_{d}}d^{d}x\sqrt{-h}(K-T)\,,\label{eq:action1}
\end{equation}
where one has the Einstein-Hilbert action with a negative cosmological constant $\Lambda<0$ in a (d+1)-dimensional spacetime $\mathcal{M}_{d+1}$ which contains a codimension-one Karch-Randall brane $\partial \mathcal{M}_{d}$ as the end-of-the-world (EoW) with tension $T$. The bulk geometry of $\mathcal{M}_{d+1}$ is determined by solving Einstein's field equations and the place of the brane can be found by solving the associated Israel's junction condition \cite{Fujita:2011fp,Takayanagi:2011zk}. 

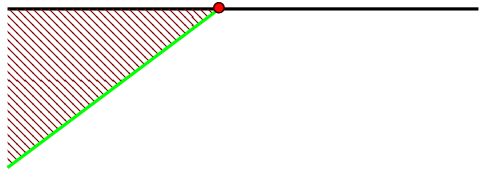
\begin{figure}[h]
\begin{center}
\begin{tikzpicture}[scale=1.4]
\draw[-,very thick,black!100] (-2,0) to (0,0);
\draw[-,very thick,black!100] (0,0) to (2.45,0);
\draw[pattern=north west lines,pattern color=purple!200,draw=none] (0,0) to (-2,-1.5) to (-2,0) to (0,0);
\draw[-,very thick,color=green!!50] (0,0) to (-2,-1.5);
\node at (0,0) {\textcolor{red}{$\bullet$}};
\node at (0,0) {\textcolor{black}{$\circ$}};
\end{tikzpicture}
\caption{\small The Karch-Randall braneworld where the bulk is empty AdS$_{d+1}$ (in Poincar\'{e} patch) and the brane (the green slice) geometry is AdS$_{d}$. The brane is an end-of-the-world so the shaded region is cut off.  This figure shows a constant-time slice.\label{figislandkr}}
\end{center}
\end{figure}

In the usual case of the Karch-Randall braneworld, the bulk geometry is asymptotic AdS$_{d+1}$ and the induced geometry on the brane is static and asymptotic AdS$_{d}$ (see Fig.~\ref{figislandkr}). The asymptotic boundary of the brane (the red defect in Fig.~\ref{figislandkr}) is a codimension-one submanifold of the asymptotic boundary of the bulk, which is dual to a conformal boundary of the dual CFT$_{d}$ of the bulk. Using AdS/CFT duality, this system has three equivalent descriptions \cite{Geng:2020qvw}:
\begin{itemize}
    \item \textbf{Bulk Description}: A (d+1)-dimensional asymptotically AdS gravitational theory with a Karch-Randall brane.
    \item \textbf{Intermediate Description}: A d-dimensional asymptotically AdS gravitational theory with cutoff conformal matter that obeys leaky boundary condition into a d-dimensional non-gravitational bath.
    \item \textbf{Boundary Description}: A d-dimensional conformal field theory with a conformal boundary, i.e. a boundary conformal field theory (BCFT$_{d}$).
\end{itemize}
We note that the conformal boundary in the boundary description is a codimension-one spatial boundary along which there is time-evolution. This usual case of the Karch-Randall braneworld has been intensively used in the computation of the Page curve of black hole radiation, the construction and study of replica wormholes and entanglement islands, and the attempt to understand the absence of global symmetries in quantum gravity \cite{Geng:2020qvw,Geng:2025rov,Geng:2024xpj,Geng:2025efs,Geng:2025gns}.

In this work, we are interested in a slightly different scenario where we expect the intermediate description to contain a closed gravitational universe with no spatial boundary. This can be realized by the following consideration \cite{Kourkoulou:2017zaj}. The conformal boundary is captured by a state $\ket{B}\rangle$ which is annihilated by the stress-energy tensor flux, i.e.
\begin{equation}
    T_{\parallel\perp}\ket{B}\rangle=0\,.
\end{equation}
This is because the Euclidean signature partition function of the CFT with a conformal boundary can be thought of as the propagation of a state $\ket{B}\rangle$ in the dual channel. Thus, one could look for the holographic dual of the state
\begin{equation}
    \ket{\Psi}=e^{-\tau H}\ket{B}\rangle\,,\label{eq:psi}
\end{equation}
with the state $\ket{B}\rangle$ defined on a closed manifold like $S^{d-1}$, for which $H$ is the CFT Hamiltonian on the cylinder $S^{d-1}\times \mathbb{R}$. As we will review in Sec.~\ref{sec:motivation}, this state is fully captured by the Karch-Randall braneworld and its intermediate description contains a closed gravitational universe. This is because now the boundary description is a CFT on a spatially closed manifold so the brane in the bulk description wouldn't be able to spatially intersect the asymptotic boundary. As a result, the brane geometry must be spatially closed as well. However, as opposed to the usual case of the Karch-Randall braneworld, the gravitational theory of the closed universe is not analyzable in this example, as the dual bulk geometry is not in a warped form. Thus, in Sec.~\ref{sec:doableexample} we will construct an analogous model which captures essential features of the above example and for which the gravitational theory on the closed universe can be explicitly analyzed as in the usual case of the Karch-Randall braneworld \cite{Karch:2000ct,Geng:2025rov}. The detailed analysis of the gravitational theory on the closed universe will be performed in Sec.~\ref{sec:observer}. 

\subsection{A Motivating Example}\label{sec:motivation}
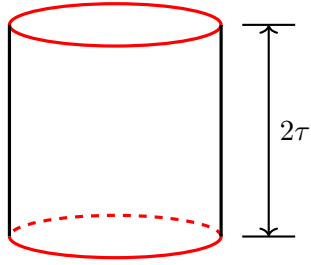
\begin{figure}[h]
\begin{center}
\begin{tikzpicture}[scale=1.4]
\draw[-,very thick,red!100] (1,1) arc (0:360:1 and 0.2);
\draw[-,very thick,dashed,red!100] (1,-1) arc (0:180:1 and 0.2);
\draw[-,very thick,red!100] (1,-1) arc (0:-180:1 and 0.2);
\draw[-,very thick,black!100] (1,1) to (1,-1);
\draw[-,very thick,black!100] (-1,1) to (-1,-1);
\draw[-,thick,black!100] (1.2,1) to (1.7,1);
\draw[-,thick,black!100] (1.2,-1) to (1.7,-1);
\draw[<->,thick,black!100] (1.45,1) to (1.45,-1);
\node at (1.7,0) {\textcolor{black}{$2\tau$}};
\end{tikzpicture}
\caption{\small The boundary configuration of the inner product Equ.~(\ref{eq:innerproduct}). The CFT path integral is computed on a cylinder with an initial state $\ket{B}\rangle$ propagating for an Euclidean time $2\tau$ and then projected to a finial state $\ket{B}\rangle$.\label{pic:CFTinnerprodcut}}
\end{center}
\end{figure}

\begin{figure}[h]
\begin{center}
\subfloat[Bulk as empty AdS$_{d+1}$ and two disconnected branes\label{pic:2branes}]
{
\begin{tikzpicture}[scale=1.4]
\draw[-,very thick,red!100] (1,1) arc (0:360:1 and 0.2);
\draw[-,very thick,dashed,red!100] (1,-1) arc (0:180:1 and 0.2);
\draw[-,very thick,red!100] (1,-1) arc (0:-180:1 and 0.2);
\draw[-,very thick,black!100] (1,1) to (1,-1);
\draw[-,very thick,black!100] (-1,1) to (-1,-1);
\draw[-,thick,black!100] (1.2,1) to (1.7,1);
\draw[-,thick,black!100] (1.2,-1) to (1.7,-1);
\draw[<->,thick,black!100] (1.45,1) to (1.45,-1);
\node at (1.7,0) {\textcolor{black}{$2\tau$}};
\draw[-,thick,red!!100] (-1,1)..controls (0,0)..(1,1);
\draw[fill=red, draw=none, fill opacity = 0.1](-1,1)..controls (0,0)..(1,1)arc (0:180:1 and 0.2);
\draw[-,thick,red!!100] (-1,-1)..controls (0,0)..(1,-1);
\draw[fill=red, draw=none, fill opacity = 0.1](-1,-1)..controls (0,0)..(1,-1)arc (0:-180:1 and 0.2);
\end{tikzpicture}}
\hspace{1cm}
\subfloat[Bulk is a black hole and only one brane\label{pic:1brane}]
{\begin{tikzpicture}[scale=1.4]
\draw[-,very thick,red!100] (1,1) arc (0:360:1 and 0.2);
\draw[-,very thick,dashed,red!100] (1,-1) arc (0:180:1 and 0.2);
\draw[-,very thick,red!100] (1,-1) arc (0:-180:1 and 0.2);
\draw[-,very thick,black!100] (1,1) to (1,-1);
\draw[-,very thick,black!100] (-1,1) to (-1,-1);
\draw[-,thick,black!100] (1.2,1) to (1.7,1);
\draw[-,thick,black!100] (1.2,-1) to (1.7,-1);
\draw[<->,thick,black!100] (1.45,1) to (1.45,-1);
\node at (1.7,0) {\textcolor{black}{$2\tau$}};
\draw[-,thick,red!!100] (-1,1)..controls (0,0)..(-1,-1);
\draw[-,thick,red!!100] (1,1)..controls (0,0)..(1,-1);
\draw[fill=red, draw=none, fill opacity = 0.1](-1,1)..controls (0,0)..(-1,-1)arc (180:360:1 and 0.2)..controls (0,0)..(1,1) arc (0:180:1 and 0.2);
\end{tikzpicture}}
\caption{\small The two possible bulk dual geometry whose on-shell action computes Equ.~(\ref{eq:innerproduct}).\label{pic:bulkinnerproduct}}
\end{center}
\end{figure}
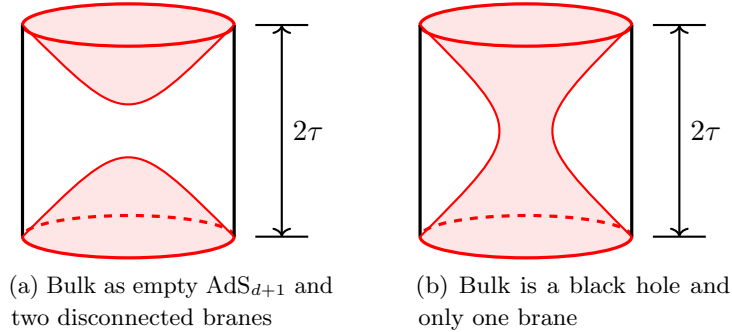

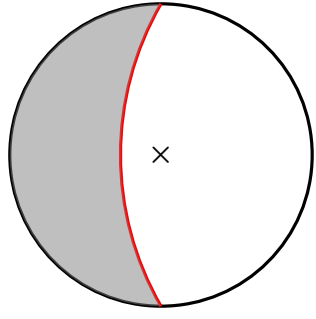
\begin{figure}
\begin{centering}
    \begin{tikzpicture}[scale=0.8]
    \draw[-,very thick,black] (2.5,0) arc (0:360:2.5);
    \draw[-,very thick,red!100] (0,2.5) arc (150:210:5);
     \draw[fill=gray, draw=none, fill opacity = 0.5] (0,2.5) arc (150:210:5) arc (-90:-270:2.5);
\node at (0,0) {\textcolor{black}{$\cross$}};
     \end{tikzpicture}
\caption{\small The bulk geometry in the $r-\tau$ plane when $d=2$. The cross denotes the black horizon which sits at the center of this Euclidean disk. The brane is the red slice behind the black hole horizon.\label{pic:Euclideanbulk}}
\end{centering}
\end{figure}

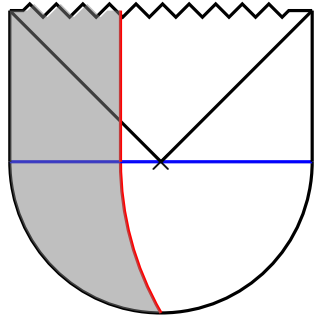
\begin{figure}
\begin{centering}
    \begin{tikzpicture}[scale=0.8]
    \draw[-,very thick,black] (2.5,0) arc (0:-180:2.5);
    \draw[-,very thick,red!100] (0,-2.5) arc (-150:-180:5);
    \draw[-,very thick,blue!100] (2.5,0) to (-2.5,0);
     \draw[fill=gray, draw=none, fill opacity = 0.5] (0,-2.5) arc (-150:-180:5)--(-2.5,0) arc (180:270:2.5);
\node at (0,0) {\textcolor{black}{$\cross$}};
\draw[-,very thick, black!100] (0,0) to (-2.5,2.5);
\draw[-,very thick, black!100] (0,0) to (2.5,2.5);
\draw[-,very thick, black!100] (-2.5,0) to (-2.5,2.5);
\draw[-,very thick, black!100] (2.5,0) to (2.5,2.5);
  \draw[-,very thick] 
       decorate[decoration={zigzag,pre=lineto,pre length=5pt,post=lineto,post length=5pt}] {(-2.5,2.5) to (2.5,2.5)};
\draw[-,very thick,red!100] (-0.66987,0)--(-0.66987,2.5);
 \draw[fill=gray, draw=none, fill opacity = 0.5] (-0.66987,0)--(-0.66987:2.5) decorate[decoration={zigzag,pre=lineto,pre length=5pt,post=lineto,post length=5pt}] {(-0.66987,2.5) to (-2.5,2.5)}--(-2.5,0)--(-0.66987,0);
     \end{tikzpicture}
\caption{\small The bulk geometry in the $r-\tau$ plane when $d=2$. We cut open the Euclidean path integral in Fig.~\ref{pic:bulkinnerproduct} at the zero-time slice (the blue slice) and turn on Lorentzian time evolution afterwards. The unshaded region of the blue slice fully captures the state $\ket{\Psi}$.\label{pic:statemotivation}}
\end{centering}
\end{figure}

\begin{figure}
\begin{centering}
    \begin{tikzpicture}[scale=0.6]
   \draw[-,very thick,black] (-1,0) arc (-180:180:1 and 0.5);
   \draw[-,very thick,dashed] (-4,-3) arc (90:270:0.5 and 1);
    \draw[-,very thick,black,dashed] (-4,-3) arc (90:-90:0.5 and 1);
    \draw[-,very thick,black] (-1,0) arc (0:-90:3);
    \draw[-,very thick,black] (1,0) arc (0:-90:5);
    \draw[-,very thick,red] (-4,-3) arc (-90:-180:1 and 3);
    \draw[-,very thick,red] (-4,-5) arc (-90:-180:3 and 5);
    \draw[-,very thick,red] (-7,0) arc (-180:180:1 and 0.5);
     \end{tikzpicture}
\caption{\small The intermediate picture path integral that prepares the entangled state between the gravitational universe (the red part) and the non-gravitational system (the black part).}\label{pic:intermediatepathintegral}
\end{centering}
\end{figure}
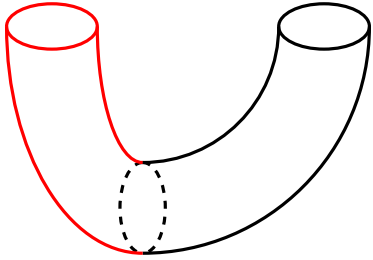

Let's consider the dual bulk geometry that computes the inner product of the state in Equ.~(\ref{eq:psi}), i.e. whose on-shell action computes
\begin{equation}
    Z=\bra{\Psi}\Psi\rangle=\langle\bra{B}e^{-2\tau H}\ket{B}\rangle\,.\label{eq:innerproduct}
\end{equation}
The boundary CFT description of this inner-product is computed on the cylinder $S^{d-1}\times\mathbb{R}_{[-\tau,\tau]}$ in Fig.~\ref{pic:CFTinnerprodcut}. As discussed in detail in \cite{Cooper:2018cmb}, there are two possible bulk duals as depicted in Fig.~\ref{pic:bulkinnerproduct}. We are interested in the second possible phase in Fig.~\ref{pic:1brane}, where the spatial geometry of the brane is $S^{d-1}$ with a nonzero size. This phase will dominate over the  phase in Fig.~\ref{pic:2branes} when the parameter $\tau>0$ is small enough. In the Euclidean signature, the bulk geometry is given by
\begin{equation}
    ds^2=f(r)d\tau^2+\frac{dr^2}{f(r)}+r^2 d\Omega_{d-1}^2\,,\label{eq:bulkmetricmotivation}
\end{equation}
where $f(r)=1+r^{2}-\frac{16\pi G_{N}M}{(d-1)\Omega_{d-1}r^{d-2}}$ with the AdS length scale set to one. In the $r-\tau$ plane the bulk geometry looks like Fig.~\ref{pic:Euclideanbulk}. If, instead of the inner-product Equ.~(\ref{eq:innerproduct}), one is interested in the state $\ket{\Psi}$ and its Lorentzian time evolution, one can cut open Fig.~\ref{pic:Euclideanbulk} at the zero-time slice and then turn on Lorentzian time evolution afterwards (see Fig.~\ref{pic:statemotivation}). In this case, the three equivalent descriptions of the system are
\begin{itemize}
    \item \textbf{Bulk Description}: A (d+1)-dimensional gravitational theory on a AdS$_{d+1}$ Schwarzschild black hole background with a Karch-Randall brane behind the black hole horizon.
    \item \textbf{Intermediate Description}: A d-dimensional closed gravitational universe with cutoff conformal matter entangled with an external non-gravitational system. This entangled state is prepared by the path integral on the manifold as depicted in Fig.~\ref{pic:intermediatepathintegral}.
    \item \textbf{Boundary Description}: A d-dimensional CFT path integral preparing the state $\ket{\Psi}$ and evaluating its norm.
\end{itemize}
Our primary interest is in the intermediate description where there is a closed gravitational universe, and we want to understand the induced gravitational theory on it and whether the Hilbert space of it is trivial. In the usual case of the Karch-Randall braneworld, this kind of analysis can be done by dimensional reduction. This is because the bulk geometry is in the warped form
\begin{equation}
    ds^2=d\rho^2+e^{2A(\rho)}ds_{d}^{2}\,,\label{eq:warpped}
\end{equation}
where $A(\rho)$ is the warped factor and the brane is one of the constant-$\rho$ slices. Thus, the perturbative aspects of the gravitational theory on the brane can be analyzed by performing the Kaluza-Klein decomposition of the bulk linearized Einstein's equation. In this analysis, one should identify the lowest KK mode of the bulk graviton, which should be localized on the brane, as the graviton in the intermediate description. However, in the current case, the bulk geometry Equ.~(\ref{eq:bulkmetricmotivation}) cannot be put in such a warped form. This obstructs an explicit analysis of the induced gravitational theory in the intermediate description. 

However, one can already deduce some properties of the induced gravitational theory in the intermediate description 
for that closed universe. For example, the graviton in the closed universe must be some combination of the modes of the bulk graviton, and such modes must all be normalizable. Due to the infinite volume of the bulk, they should all be massive, as the massless mode cannot be normalizable. This is a basic property of the Karch-Randall braneworld and has shed light on various puzzles in quantum gravity. Thus, to have a better understanding of the physics of the closed universe, we would like to have a model, which captures the essence of the motivating example we considered in this subsection but with an explicit analysis doable. We will consider such a model in the next subsection, and this model is motivated by the Maldacena-Maoz wormhole \cite{Maldacena:2004rf}.

\subsection{An Analyzable Model}\label{sec:doableexample}
The model we are now considering has the bulk geometry
\begin{equation}
    ds^2=d\rho^2+\cosh^{2}\rho (-dt^2+\cos^2 td\Sigma_{d-1}^2)\,,\label{eq:bulkmetric}
\end{equation}
where $\rho\in(-\infty,\infty)$ when there is no brane and $\Sigma_{d-1}$ is a closed hyperbolic manifold that is constructed as the quotient of a $(d-1)$-dimensional Euclidean empty AdS geometry by a discrete subgroup $\Gamma$ of its isometry group $SO(d-1,1)$, i.e.
\begin{equation}
    \Sigma_{d-1}=\text{EAdS}_{d-1}/\Gamma\,.
\end{equation}
This bulk geometry solves Einstein's equation with a negative cosmological constant and is actually a smooth quotient of the empty AdS$_{d+1}$. We put one Karch-Randall brane in this bulk at $\rho=\rho_{B}$, where the brane tension is
\begin{equation}
    T=(d-1)|\tanh\rho_{B}|\,,\label{eq:tension}
\end{equation}
to model the closed gravitational universe as in Sec.~\ref{sec:motivation}. The induced geometry on the brane is
\begin{equation}
    ds_{\text{brane}}^2=-dt^2+\cos^2 t d\Sigma_{d-1}^{2}\,,\label{eq:braneFRW}
\end{equation}
whose spatial section is the closed manifold $\Sigma_{d-1}$. 

The bulk geometry Equ.~(\ref{eq:bulkmetric}) is in a warped form with the volume infinite ($\cosh\rho$ goes to infinity exponentially fast as $\rho\rightarrow\pm\infty$). This allows us to perform explicit analysis of the induced gravitational theory on the closed universe as in the usual case of the Karch-Randall braneworld. The analysis and results will be listed in Sec.~\ref{sec:observer}. In the rest of this subsection, let's work out a few geometric aspects of this model to articulate its three equivalent descriptions.

In fact, the metric Equ.~(\ref{eq:bulkmetric}) describes the hyperbolic black hole in AdS$_{d+1}$ \cite{Emparan:1999gf}. This connection manifests itself through the following coordinates transform
\begin{equation}
    r=\cosh\rho\cos t\,,\quad \tanh\eta=\coth\rho\sin t\,,
\end{equation}
by which the metric becomes
\begin{equation}
    ds^2=-(r^2-1)d\eta^2+\frac{dr^2}{r^2-1}+r^2 d\Sigma_{d-1}^{2}\,,
\end{equation}
and we have a horizon at $r=1$. Using Equ.~(\ref{eq:tension}), we can see that in this coordinate the brane is described by the following hypersurface
\begin{equation}
    r^2 ((d-1)^2-T^2)+T^2 \tanh^2\eta=(d-1)^2\,,
\end{equation}
where we only consider positive tension branes with the bulk asymptotic boundary sitting at $r\rightarrow+\infty$. The state interpretation of this setup in the boundary description can be understood in the Euclidean signature with $\tilde\eta=i\eta$ and
\begin{equation}
    ds^2=(r^2-1)d\tilde\eta^2+\frac{dr^2}{r^2-1}+r^2 d\Sigma_{d-1}^{2}\,.\label{eq:bulkhbh}
\end{equation}
We have $\tilde\eta\in[0,2\pi]$ and $r\in[1,\infty)$, so in the $r-\tilde\eta$ plane the above Euclidean geometry is a disk with the center at $r=1$. The brane is now described by
\begin{equation}
        r^2 ((d-1)^2-T^2)-T^2 \tan^2\tilde\eta=(d-1)^2\,,
\end{equation}
so the brane hits the asymptotic boundary at $\tilde\eta=\pm\frac{\pi}{2}$. Thus, the configuration of our setup in the $r-\tilde\eta$ plane is the same as Fig.~\ref{pic:Euclideanbulk} for any dimension $d$, and the situation in the Lorentzian signature is almost the same as depicted in Fig.~\ref{pic:statemotivation}. As a result, the boundary description of this model is that the $t=\eta=0$ slice is described by the CFT state
\begin{equation}
    \ket{\Phi}=e^{-\frac{\pi}{2}H_{\eta}}\ket{B}\rangle\,,
\end{equation}
prepared on the Euclidean cylinder
\begin{equation}
    ds^2_{\text{CFT}}=d\tilde\eta^2+d\Sigma_{d-1}^2\,,
\end{equation}
with a boundary state $\ket{B}\rangle$, that corresponds to the Karch-Randall brane with tension $T$, starting at $\tilde\eta=-\frac{\pi}{2}$, and this state is evolved by the CFT Hamiltonian $H_{\eta}$ on the above cylinder to $\tilde\eta=0$. Therefore, the three equivalent descriptions of this model are in the same form as the motivating example in Sec.~\ref{sec:motivation}. We are interested in understanding the physics on the closed gravitational universe in the intermediate description.

\section{Uncovering the Observer and the Nontrivial Hilbert Space}\label{sec:observer}
In Sec.~\ref{sec:Construction}, we have seen that a closed gravitational universe can be naturally realized in the Karch-Randall braneworld and we provided an analyzable model in Sec.~\ref{sec:doableexample}. This model is closely related to the models designed to understand the physics of black hole radiation \cite{Geng:2024xpj} and in some sense the closed universe in the intermediate description can be thought of as the black hole interior \cite{Kourkoulou:2017zaj,Cooper:2018cmb,Penington:2019kki,Antonini:2021xar,Antonini:2024bbm,Geng:2024jmm}. Hence, we believe that this model will be illuminating in generalizing and refining the lessons we learned about black holes \cite{Geng:2020qvw,Geng:2020fxl,Geng:2021hlu,Geng:2024jmm,Geng:2024xpj,Geng:2025rov,Geng:2025efs,Geng:2025gns,Geng:2025byh,Geng:2026asi} in this new context of closed universes. This is both interesting and important, as our current universe has a positive cosmological constant \cite{Planck:2018vyg}, and the simplest solution of Einstein's equation with such a feature is spatially closed. 

In this section, we will perform a preliminary analysis of the model we provided in Sec.~\ref{sec:doableexample}. We will see that the Hilbert space associated with the closed gravitational universe is necessarily nontrivial due to the existence of ``observer" in the closed universe. We will comment on the nature of this ``observer", its relationship with other proposals in the literature and its implications in Sec.~\ref{sec:comments}. We will perform the analysis using two parallel approaches---from the bulk description and directly within the intermediate description. We find consistent results in these two approaches.

\subsection{The Absence of Isometries and Perturbatively Localized Observables}\label{sec:noisometry}
Before we delve into the details of our system, let's make a few remarks regarding the closed universe Equ.~(\ref{eq:braneFRW}) and the implication of the argument for the trivial Hilbert space. We will first show that there are no isometries in the closed universe Equ.~(\ref{eq:braneFRW}) and this allows us to construct perturbatively localized observables inside this closed gravitational universe. We will argue that the argument for the trivial Hilbert space will lift these observables. Thus, even though they are interesting, these observables may not be relevant to address or relax the argument for the trivial Hilbert space. This motivates us to consider observables in Sec.~\ref{sec:bulk} and Sec.~\ref{sec:intermediate}.

Let's denote the background metric on the closed universe Equ.~(\ref{eq:braneFRW}) as
\begin{equation}
    ds^{2}=-dt^2+\cos^2t d\Sigma_{d-1}^2=\bar{g}_{ij}dx^{i}dx^{j}\,.
\end{equation}

First, we notice that there is no continuous isometry of the closed universe Equ.~(\ref{eq:braneFRW}). Let's prove this by contradiction. Assume that there is a globally defined Killing vector field
\begin{equation}
    U(x)=\gamma(x)\partial_{t}+u(x)\,,
\end{equation}
where $u(x)$ is tangential to the closed spatial slice $\Sigma_{d-1}$. The Killing equation reads
\begin{equation}
    \bar{\nabla}_{i}U_{j}+\bar{\nabla}_{j}U_{i}=0\,,
\end{equation}
and it implies
\begin{equation}
    \partial_{t}\gamma=0\,,\quad \cos^2t\partial_{t}\hat{u}_{a}-\partial_{a}\gamma=0\,,\quad \cos^2t(\hat{\nabla}_{a}\hat{u}_{b}+\hat{\nabla}_{b}\hat{u}_{a})-2\gamma\cos t\sin t\hat{g}_{ab}=0\,,\label{eq:Killingequations}
\end{equation}
where we denote the metric on $\Sigma_{d-1}$ as $\hat{g}_{ab}$, $\hat{u}_{a}=\hat{g}_{ab}u^{b}$ and $\hat{\nabla}_{a}$'s are the metric compatible and torsion free covariant derivatives with respect to $\hat{g}_{ab}$. The above equations must be satisfied for any $t$. Thus, we can first consider the $t=0$ slice. From the third equation of Equ.~(\ref{eq:Killingequations}), we have 
\begin{equation}
(\hat{\nabla}_{a}\hat{u}_{b}+\hat{\nabla}_{b}\hat{u}_{a})\big\vert_{t=0}=0\,,\label{eq:u}
\end{equation}
and by combining the second equation and the $t$-derivative of the third equation, we obtain\footnote{Note that the first equation in Equ.~(\ref{eq:Killingequations}) implies that $\gamma$ is independent of $t$.}
\begin{equation}
    (\hat{\nabla}_{a}\partial_{b}+\hat{\nabla}_{b}\partial_{a})\gamma=2\gamma \hat{g}_{ab}\,.\label{eq:gamma}
\end{equation}
From Equ.~(\ref{eq:gamma}) and the compactness of $\Sigma_{d-1}$ we have
\begin{equation}
    -\int_{\Sigma_{d-1}}d^{d-1}x\sqrt{\hat{g}}\hat{g}^{ab} \partial_{a}\gamma\partial_{b}\gamma=(d-1)\int_{\Sigma_{d-1}} d^{d-1}x\sqrt{\hat{g}} \gamma^2\,,
\end{equation}
which implies that
\begin{equation}
    \gamma=0\,,\quad\text{for $d>1$}\,.
\end{equation}
From the second equation of Equ.~(\ref{eq:Killingequations}) this further implies
\begin{equation}
    \partial_{t}\hat{u}_{a}=0\,,
\end{equation}
and so we can remove the restriction to $t=0$ in Equ.~(\ref{eq:u}). As a result, we have
\begin{equation}
    \hat{\nabla}^{a}\hat{\nabla}_{a}\hat{u}_{b}=-\hat{\nabla}^{a}\hat{\nabla}_{b}\hat{u}_{a}=-\hat{R}^{a}_{b}\hat{u}_{a}-\hat{\nabla}_{b}\hat{\nabla}^{a}\hat{u}_{a}=-\hat{R}^{a}_{b}\hat{u}_{a}=(d-2)\hat{u}_{b}\,.
\end{equation}
Again, using the fact that $\Sigma_{d-1}$ is compact, we have
\begin{equation}
    -\int_{\Sigma_{d-1}}d^{d-1}x\sqrt{\hat{g}}\hat{g}^{ab} \hat{g}^{cd} \hat{\nabla}_{a}\hat{u}_{c}\hat{\nabla}_{b}\hat{u}_{d}=(d-2)\int_{\Sigma_{d-1}} d^{d-1}x\sqrt{\hat{g}} \hat{g}^{ab}\hat{u}_{a}\hat{u}_{b}\,,
\end{equation}
which implies that
\begin{equation}
    \hat{u}_{a}=0\,,\quad\text{for $d>2$}\,.
\end{equation}
In summary, we have 
\begin{equation}
    U(x)=0\,,\quad\text{for $d>2$}.
\end{equation}
Thus, there is no Killing vector for the higher dimensional cases we are interested in.

The above result for the absence of a Killing vector in the closed gravitational universe we are interested in suggests that the diffeomorphisms are already spontaneously broken by the background geometry itself. Thus, following the general philosophy of \cite{Geng:2024dbl}, one should be able to construct diffeomorphism invariant observables. Since this particular type of example was not explicitly considered in \cite{Geng:2024dbl}, let's first provide an explicit construction of diffeomorphism invariant observables dressed to the background geometric features that spontaneously break the diffeomorphisms. The goal is to construct a Goldstone vector boson field $v^{i}$ that transforms nonlinearly under the diffeomorphisms as \footnote{We note that we shouldn't confuse this Goldstone vector boson with the Goldstone vector boson we will discuss in Sec.~\ref{sec:inducedtheory} and the later sections of this paper. Thus, we use $v^{i}(x)$ here.}
\begin{equation}
    v^{i}(x)\rightarrow v^{i}(x)-\epsilon^{i}(x)\,.
\end{equation}
As proposed in \cite{Geng:2024dbl}, this vector field should be constructed from the fluctuations of the symmetry breaking background features, i.e. the background metric in the current case. Let's consider the linearized graviton field $h_{ij}$ on this closed universe, and we know that it transforms under the diffeomorphisms as
\begin{equation}
    h_{ij}(x)\rightarrow h_{ij}(x)+\bar{\nabla}_{i}\epsilon_{j}(x)+\bar{\nabla}_{j}\epsilon_{i}(x)\,,\quad\text{with }x^{i}\rightarrow x^{i}+\sqrt{16\pi G_{N}}\epsilon^{i}(x)
\end{equation}
Let's define
\begin{equation}
    C^{i}(x)=\bar{g}^{ij}(\bar{\nabla}^{k}h_{kj}-\frac{1}{2}\bar{\nabla}_{j}h)(x)\,.
\end{equation}
Thus, we have the transformation under the diffeomorphisms
\begin{equation}
    C^{i}(x)\rightarrow C^{i}(x)+\bar{\nabla}^{2}\epsilon^{i}(x)-(d-1)\epsilon^{i}(x)\,.
\end{equation}
Hence, we have
\begin{equation}
    v^{i}(x)=-\Big(\big(\bar{\nabla}^{2}-(d-1)\big)^{-1}C\Big)^{i}(x)\,,
\end{equation}
which transforms under the diffeomorphisms indeed as
\begin{equation}
    v^{i}(x)\rightarrow v^{i}(x)-\epsilon^{i}(x)\,.
\end{equation}
We note that inverse differential operator $\big(\bar{\nabla}^{2}-(d-1)\big)^{-1i}_{k}$ is simply the Green's function of a massive vector field in the closed universe Equ.~(\ref{eq:braneFRW}). With this Goldstone vector boson obtained, we can dress probe matter operators in a diffeomorphism invariant manner. For example, for a scalar probe operator $\hat{O}(x)$, we have
\begin{equation}
    \hat{O}^{\text{Phys}}(x)=\hat{O}(x+\sqrt{16\pi G_{N}}\hat{v}(x))\,,\label{eq:Opert}
\end{equation}
where we used $\hat{}$ to denote operators. As discussed in \cite{Geng:2024dbl} and \cite{Geng:2025gqu}, these dressed operators are diffeomorphism invariant and are perturbatively localized operators that cannot be detected in any nonlocal manner.\footnote{ This dressing construction can be improved and promoted to any order in gravitational perturbation theory.} As a consistency check, we can see that when there are Killing vectors for the background metric, $C^{i}(x)$ is invariant under the Killing part of the diffeomorphism, i.e., the isometry, so $v^{i}(x)$ would also be invariant. Thus, when there exist Killing vectors, the above constructed observables fail to be diffeomorphism invariant. 

For our later purpose, it is helpful to see explicitly how the observables in the form of Equ.~(\ref{eq:Opert}) obey the Hamiltonian constraint. Let's quickly setup the framework for this analysis. The ADM decomposition of the metric \cite{Arnowitt:1962hi} on a d-dimensional spacetime is
\begin{equation}
ds^{2}=-N^{2} dt^{2}+g_{ab}(dx^{a}+N^{a}dt)(dx^{b}+N^{b}dt)\,,\label{eq:ADM}
\end{equation}
where $N$ is called the \textit{lapse function}, the vector $N^{a}$ is called the \textit{shift vector}. This particular decomposition can be thought of as a general gauge fixing procedure that fixes the $d$-coordinate reparametrizations. We can write the matter-coupled Einstein-Hilbert action, using the Gauss-Codazzi equation, as
\begin{equation}
S=\frac{1}{16\pi G_{N}}\int dt d^{d-1}x N\sqrt{g}(R[g]-2\Lambda+K_{ab}K^{ab}-K^{2})+S_{\text{matter}}+S_{\text{bdy}}\,,\label{eq:actionADM}
\end{equation}
where we have the cosmological constant $\Lambda=-\frac{(d-1)(d-2)}{2}$, the Ricci scalar of the spatial metric $g_{ab}$ as $R[g]$, the action of the minimally coupled matter fields $S_{\text{matter}}$, the appropriate boundary term of the action $S_{\text{bdy}}$ to ensure a well-defined variational principle, and the extrinsic curvature of the constant-$t$ slices $K_{ab}$. The extrinsic curvature can be computed from Equ.~(\ref{eq:ADM}) as
\begin{equation}
K_{ab}=\frac{1}{2N}(-\dot{g}_{ab}+\nabla_{a}N_{b}+\nabla_{b}N_{a})\,,\label{eq:K}
\end{equation}
where $\nabla^a$ is the metric compatible and torsion free covariant derivative with respect to the spatial metric $g_{ab}$. From the action Equ.~(\ref{eq:actionADM}) and the extrinsic curvature Equ.~(\ref{eq:K}), it is easy to see that the canonical momenta associated with the lapse function and shift vector are zero
\begin{equation}
\Pi=\frac{1}{\sqrt{g}}\frac{\delta S}{\delta \dot{N}}=0\,,\quad \Pi_{a}=\frac{1}{\sqrt{g}}\frac{\delta S}{\delta \dot{N^{a}}}=0\,.\label{eq:primary}
\end{equation}
We can obtain the Hamiltonian of the system by doing the Legendre transform, and the result is
\begin{equation}
H_{\text{tot}}=\int d^{d-1}x\sqrt{g}\Big[N\mathcal{H}+N^{a}\mathcal{H}_{a}\Big]+H_{\text{bdy}}\,,\label{eq:Htot}
\end{equation}
where
\begin{equation}
\begin{split}
  \mathcal{H}&=16\pi G_{N}\Big(\Pi_{ab}\Pi^{ab}-\frac{1}{d-2}(\Pi^{a}_{a})^{2}\Big)-\frac{1}{16\pi G_{N}}(R[g]-2\Lambda)+\mathcal{H}_{\text{matter}}\,,\\
\mathcal{H}_{a}&=-2g_{ab}\nabla_{c}\Pi^{bc}+\mathcal{H}_{a,\text{matter}}\,.\label{eq:constraints}
  \end{split}
\end{equation}
In the above formula, we denoted  the Hamiltonian density of the matter field as $\mathcal{H}_{\text{matter}}$, $\mathcal{H}_{a,\text{matter}}$'s are the momentum densities of the matter fields, and we have the canonical momentum of $g_{ab}$
\begin{equation}
\Pi^{ab}=\frac{1}{\sqrt{g}}\frac{\delta S}{\delta \dot{g}_{ab}}=-\frac{1}{16\pi G_{N}}\Big(K^{ab}-g^{ab}K\Big)\,.\label{eq:Pigraviton}
\end{equation}
The equations in Equ.~(\ref{eq:primary}) are called \textit{primary constraints} \cite{dirac2001quantum}, and they must be preserved under the time-evolution generated by the Hamiltonian in Equ.~(\ref{eq:Htot}). From this condition, we have \textit{secondary constraints}
\begin{equation}
\mathcal{H}=0\,,\quad \mathcal{H}_{i}=0\,.
\end{equation}
These secondary constraints constrain physical states and observables in the quantum theory, i.e. after we promote $\mathcal{H}$ and $\mathcal{H}_{i}$ to operators $\hat{\mathcal{H}}$ and $\hat{\mathcal{H}}_{i}$. More explicitly, physical states in the Hilbert space of the system have to be annihilated by $\hat{\mathcal{H}}$ and $\hat{\mathcal{H}}_{i}$, and meanwhile physical observables have to commute with them.

In this paper, we are primarily interested in the constraint $\mathcal{H}=0$ which is called the \textit{Hamiltonian constraint}. We will do the perturbative analysis around the background metric Equ.~(\ref{eq:braneFRW}), in which $N=1$ and  $N_{a}=0$, and the background metric $g_{ab}^{0}$ is given by
\begin{equation}
g_{ab}^{0}dx^{a}dx^{b}=\cos^2 t d\Sigma_{d-1}^2=\cos^2 t \hat{g}_{ab}dx^a dx^b\,.\label{eq:bc}
\end{equation}
We have the background Ricci scalar and the extrinsic curvature
\begin{equation}
    R[g^{0}]=-\frac{1}{\cos^2 t}(d-2)(d-1)\,, K_{ab}^{0}=\cos t\sin t\hat{g}_{ab}\,.\label{eq:backgroundKandR}
\end{equation}
Because the lapse $N$ and shift $N_{i}$ are constrained to be nondynamical, only the spatial metric $g_{ab}$ is dynamical. Since we are doing perturbative theory, we can expand the spatial metric as
\begin{equation}
g_{ab}=g_{ab}^{0}+\sqrt{16\pi G_{N}}h_{ab}\,,
\end{equation}
where we only consider the leading nontrivial order perturbative effects and $h_{ab}$ is the graviton field which has a normalized kinetic term from expanding the Einstein-Hilbert action to quadratic order in this graviton field. To the leading order in $G_{N}$ we have 
\begin{equation}
\mathcal{H}^{(0)}=-\frac{1}{16\pi G_{N}}(R[g^{0}]-2\Lambda-K^{0}_{ab}K^{0ab}+K^{02})=0\,,
\end{equation}
which can be seen using Equ.~(\ref{eq:backgroundKandR}). Thus, the Hamiltonian constraint is nontrivial starting from the first order, for which we have
\begin{equation}
\begin{split}
\mathcal{H}^{(1)}&=-\frac{1}{\sqrt{16\pi G_{N}}}\Big[\frac{(d-2)h}{\cos^2 t}+\nabla_{a}\nabla_{b}h^{ab}-\nabla^{2}h\Big]+\mathcal{H}_{\text{matter}}+2\sqrt{16\pi G_{N}}\Big(\Pi^{0}_{ab}\delta\Pi^{ab}-\frac{1}{d-2}\Pi^{0a}_{a}g^{0}_{bc}\delta\Pi^{bc}\Big)\,,\\&+\sqrt{16\pi G_{N}}\Big(\Pi^{0ab}\Pi^{0cd}(g^{0}_{ac}h_{bd}+g^{0}_{bd}h_{ac})-\frac{1}{d-2}\Pi^{0ab}\Pi^{0cd}(g^{0}_{cd}h_{ab}+g^{0}_{ab}h_{cd})\Big)\,,\label{eq:core}
\end{split}
\end{equation}
where $\delta\Pi^{ab}$ are the linearizations of the canonical momenta $\Pi^{ab}$ and in terms of the canonical conjugate momenta of the linearized graviton field and the linearized graviton field itself we have
\begin{equation}
    \delta\Pi^{ab}=\pi^{ab}-\frac{d-2}{2}g^{0ab}h\tan t\,.
\end{equation}
We have the canonical commutation relation
\begin{equation}
    [\pi^{ab}(t,\vec{x}),h_{cb}(t,\vec{y})]=-i\delta^{(a}_{(c}\delta^{b)}_{d)}\frac{1}{\sqrt{g^{0}}N}\delta^{d-1}(\vec{x}-\vec{y})\,.\label{eq:gravitoncanonical}
\end{equation}
The background canonical momenta can be computed from Equ.~(\ref{eq:Pigraviton}) and Equ.~(\ref{eq:backgroundKandR}) as
\begin{equation}
    \Pi^{0}_{ab}=\frac{d-2}{16\pi G_{N}}\tan tg^{0}_{ab}\,.
\end{equation}
Hence, we can simplify Equ.~(\ref{eq:core}) as 
\begin{equation}
\begin{split}
\mathcal{H}^{(1)}&=-\frac{1}{\sqrt{16\pi G_{N}}}\Big[\frac{(d-2)h}{\cos^2 t}+\nabla_{a}\nabla_{b}h^{ab}-\nabla^{2}h\Big]+\mathcal{H}_{\text{matter}}\\&+\frac{1}{\sqrt{16\pi G_{N}}}\Big(-2\pi^{a}_{a}\tan t+(d-2)(d-3)h\tan t\Big)\,.\label{eq:core1}
\end{split}
\end{equation}
Imposing the constraint, i.e. setting $\mathcal{H}^{(1)}=0$, we have
\begin{equation}
\begin{split}
\sqrt{16\pi G_{N}}H_{\text{matter}}&=\sqrt{16\pi G_{N}}\int d^{d}x
\sqrt{g^{0}}N\mathcal{H}_{\text{matter}}\\&=\int d^{d}x
\sqrt{g^{0}}N\Big[\frac{(d-2)h}{\cos^2t}+\nabla_{a}\nabla_{b}h^{ab}-\nabla^{2}h+2\pi^{a}_{a}\tan t-(d-2)(d-3)h\tan^2 t\Big]\,,\\&=\int d^{d}x
\sqrt{g^{0}}\Big[\frac{(d-2)h}{\cos^2t}+2\pi^{a}_{a}\tan t-(d-2)(d-3)h\tan^2 t\Big]\,,\\&\equiv H_{\text{bulk}}[h]\,,\label{eq:HADM}
\end{split}
\end{equation}
where we dropped the total derivative term as the spatial manifold $\Sigma_{d-1}$ is compact with no boundary. For future reference, we have the trace of the canonical momenta
\begin{equation}
    \pi^{a}_{a}=\frac{d-2}{2}\Big((d-3)h\tan t-\partial_{t}h\Big)\,.
\end{equation}
To check if Equ.~(\ref{eq:Opert}) obeys the Hamiltonian constraint, we have to compute the following equal-time commutator
\begin{equation}
    [\sqrt{16\pi G_{N}}H_{\text{matter}}-H_{\text{bulk}}[h],\hat{O}^{\text{Phys}}(x)]\,,
\end{equation}
and see whether it is zero. For this purpose, we can first compute the equal-time commutator
\begin{equation}
    [H_{\text{bulk}}[h],C^{i}(x)]\,.
\end{equation}
In terms of components we have
\begin{equation}
    C^{t}(x)=-\frac{\pi^{a}_{a}}{d-2}+\frac{d-5}{2}h\tan t\,,\quad C^{a}=\nabla^{b}h^{a}_{b}-\frac{1}{2}\partial_{a}h\,.
\end{equation}
It is easy to obtain the following equal-time commutators
\begin{equation}
    [H_{\text{bulk}}[h],C^{t}(x)]=i(d-1)(\tan^2t-1)\,,\quad [H_{\text{bulk}},C^{a}(x)]=0\,.
\end{equation}
As a result, we have the equal-time commutator
\begin{equation}
    [H_{\text{bulk}}[h],v^{t}(x)]=-i\Big(\big(\bar{\nabla}^{2}-(d-1)\big)^{-1}(d-1)(\tan^2t-1)\beta\Big)^{t}(x)\,,\quad[H_{\text{bulk}}[h],v^{a}(x)]=0\,,
\end{equation}
where we have defined a vector $\beta^{i}=\delta^{i}_{t}$. It is easy to see that we have
\begin{equation}
    (d-1)(\tan^2t-1)\beta^{\mu}=[\bar{\nabla}^2-(d-1)]\beta^{\mu}\,.
\end{equation}
Therefore, we have
\begin{equation}
    [H_{\text{bulk}}[h],v^{t}(x)]=-i\,,\quad[H_{\text{bulk}}[h],v^{a}(x)]=0\,.
\end{equation}
As a result, 
\begin{equation}
    [\sqrt{16\pi G_{N}}H_{\text{matter}}-H_{\text{bulk}}[h],\hat{O}^{\text{Phys}}(x)]=0\,.
\end{equation}
Thus, the Hamiltonian constraint is obeyed by the localized observables Equ.~(\ref{eq:Opert})!

Let's make a few remarks about the above results. We first note that the above construction is only to the first nontrivial order in gravitational perturbation theory, and it can be extended to all orders. The operators in the form of Equ.~(\ref{eq:Opert}) are thus perturbatively localized, and they are defined to describe low-energy excitations around the background geometry Equ.~(\ref{eq:braneFRW}). Therefore, their nonperturbative fate is not so clear. If one takes the argument for the trivial Hilbert space seriously as a nonperturbative argument, these operators would then be lifted. Therefore, they cannot be directly used to address or relax the argument for the trivial Hilbert space in a closed gravitational universe.  Nevertheless, they are not totally meaningless, as they suggest that there should be some nonperturbative signature for the operators that can be used to address the argument for the trivial Hilbert space and see how it could be relaxed. We will construct such operators and discuss their implications in the next few sections.

\subsection{From the Bulk Description}\label{sec:bulk}
As we discussed in Sec.~\ref{sec:Construction}, the warped form of the metric Equ.~(\ref{eq:bulkmetric}) allows us to understand the gravitational physics on the closed universe, i.e. the Karch-Randall brane, by the Kaluza-Klein analysis of the linearized Einstein's equation of the bulk. This analysis has been performed in detail for general cases in \cite{Geng:2025rov,Geng:2026wok}. We first list the results and their implications, and we refer the readers to \cite{Geng:2025rov} for further details and discussions.

\subsubsection{The Induced Gravitational Theory on the Brane}\label{sec:inducedtheory}
Let's first understand the induced gravitational theory in the closed universe. We expand the bulk geometry Equ.~(\ref{eq:warpped}) to components and again we use the abstract symbol $\bar{g}_{ij}(x)$ to denote the metric of the closed universe
\begin{equation}
ds^{2}=d\rho^{2}+e^{2A(\rho)}\bar{g}_{ij}(x)dx^{i}dx^{j}\,,\label{eq:warppedmetric}
\end{equation}
and in our case in Equ.~(\ref{eq:bulkmetric}) we have $e^{A(\rho)}=\cosh\rho$. We remind ourselves that the Karch-Randall brane is along a constant-$\rho$ slice $\rho=\rho_{B}$. As we discussed in Sec.~\ref{sec:doableexample}, $\rho_{B}$ is determined by the brane tension $T$ as in Equ.~(\ref{eq:tension}) and $\rho\in[\rho_{B},\infty)$ for the bulk. The linearized Einstein's equation in the AdS$_{d+1}$ bulk is
\begin{equation}
    \mathcal{E}^{(d+1)}_{\mu\nu\alpha\beta}h^{\alpha\beta}\equiv \frac{1}{2}g_{\mu\nu}\nabla_{\alpha}\nabla_{\beta}h^{\alpha\beta}-\frac{1}{2}g_{\mu\nu}\Box h+\frac{1}{2}\nabla_{\mu}\nabla_{\nu}h-\frac{1}{2}\nabla_{\alpha}\nabla_{\mu}h_{\nu}^{\alpha}-\frac{1}{2}\nabla_{\alpha}\nabla_{\nu}h_{\mu}^{\alpha}+\frac{1}{2}\Box h_{\mu\nu}+\frac{d}{2}(g_{\mu\nu}h-2h_{\mu\nu})=0\,,\label{eq:linearizedEF}
\end{equation}
where we use $g_{\mu\nu}$ to denote the background metric of the bulk. The above equation is the equation of motion of the graviton field $h_{\mu\nu}$, which is the linearization of the full dynamical metric $\hat{g}_{\mu\nu}$ around the background $g_{\mu\nu}$
\begin{equation}
    \hat{g}_{\mu\nu}=g_{\mu\nu}+\sqrt{16\pi G_{N}}h_{\mu\nu}\,.
\end{equation}
The upper indices in Equ.~(\ref{eq:linearizedEF}) come from contracting the lower indices using the inverse background metric $g^{\mu\nu}$. If the bulk background metric $g_{\mu\nu}$ obeys the vacuum Einstein's field equation, the linearized equation Equ.~(\ref{eq:linearizedEF}) is invariant under the linearized bulk diffeomorphism transform:
\begin{equation}
    h_{\mu\nu}(x,\rho)\rightarrow h_{\mu\nu}(x,\rho)+\nabla_{\mu}\epsilon_{\nu}(x,\rho)+\nabla_{\nu}\epsilon_{\mu}(x,\rho)\,,\label{eq:diffeo}
\end{equation}
where the diffeomorphism parameter $\epsilon_{\mu}$ denotes the infinitesimal diffeomorphism transform
\begin{equation}
    x^{\mu}\rightarrow x^{\mu}+\sqrt{16\pi G_{N}}\epsilon^{\mu}(x,\rho)\,.
\end{equation}
The linearized bulk Einstein's equation Equ.~(\ref{eq:linearizedEF}) can be decomposed into a set of equations with a clear $d$-dimensional interpretation. Before the decomposition, let's first define the vector field $V_{\mu}(x,\rho)$ whose components are
\begin{equation}
    \begin{split}
   V_{\rho}(x,\rho)&=\frac{1}{2}\int_{\rho}^{\infty} du h_{\rho\rho}(x,u)\,, \\ V_{i}(x,\rho)&=e^{2A(\rho)}\int_{\rho}^{\infty} du e^{-2A(u)}h_{\rho i}(u,x)+\frac{e^{2A(\rho)}}{2}\int _{\rho}^{\infty} due^{-2A(u)}\int_{u}^{\infty} du' \partial_{i}h_{\rho\rho}(x,u')\,.\label{eq:Vdef}
    \end{split}
\end{equation}
This vector field can be obtained by solving the following equation
\begin{equation}
    h_{\rho\mu}(x,\rho)=-\nabla_{\rho}V_{\mu}(x,\rho)-\nabla_{\mu}V_{\rho}(x,\rho)\,,\label{eq:hV}
\end{equation}
and it is the gravitational Wilson line in the $\rho$-direction \cite{Donnelly:2015hta,Donnelly:2016rvo,Giddings:2018umg,Giddings:2019hjc}. According to Equ.~(\ref{eq:hV}) and Equ.~(\ref{eq:diffeo}), this vector field transforms under the linearized diffeomorphism as
\begin{equation}
    V_{\mu}(x,\rho)\rightarrow V_{\mu}(x,\rho)-\epsilon_{\mu}(x,\rho)\,.\label{eq:Vtrans}
\end{equation}
Let's further define $\tilde{h}_{ij}$ as
\begin{equation}
    \bar{h}_{ij}(x,\rho)= h_{ij}(x,\rho)+2e^{2A(\rho)}A'(\rho)\bar{g}_{ij}(x)V_{\rho}(x,\rho)\,.\label{eq:newh}
\end{equation}
This new field is invariant under the linearized diffeomorphism transform Equ.~(\ref{eq:diffeo}) in the $\rho$-direction. Moreover, $\tilde{h}_{ij}$ transforms under the $i$-direction diffeomorphisms as
\begin{equation}
    \bar{h}_{ij}(x,\rho)\rightarrow \bar{h}_{ij}(x,\rho)+\bar{\nabla}_{i}\epsilon_{j}(x,\rho)+\bar{\nabla}_{j}\epsilon_{i}(x,\rho)\,,
\end{equation}
where $\bar{\nabla}_{i}$ is the metric compatible and torsion-free covariant derivative of the geometry of the closed universe $\bar{g}_{ij}(x)$.
Thus, this new field $\tilde{h}_{ij}$ is more like a ``$d$-dimensional graviton" field than the old field $h_{ij}$. Using the new field $\tilde{h}_{ij}$ and the vector field $V_{\mu}$, the bulk linearized Einstein's equation can be decomposed as the following three equations
\begin{equation}
    \begin{split}
        0=&\mathcal{E}^{(d)}_{ijlm}\bar{h}^{lm}+\frac{1}{2}e^{-(d-4)A(\rho)}\partial_{\rho}\Big[e^{dA(\rho)}\partial_{\rho}e^{-2A(\rho)}(\bar{h}_{ij}-\bar{g}_{ij}\bar{h})\Big]\\&+\frac{1}{2}e^{-(d-4)A(\rho)}\partial_{\rho}\Big[e^{dA(\rho)}\partial_{\rho}e^{-2A(\rho)}(\bar{\nabla}_{i}V_{j}+\bar{\nabla}_{j}V_{i}-\bar{g}_{ij}2\bar{\nabla}_{i}V^{i})\Big]\,,\\0=&
        \frac{1}{2}e^{-(d-4)A(\rho)}\partial_{\rho}\Big[e^{dA(\rho)}\partial_{\rho}e^{-2A(\rho)}\big[\bar{\nabla}^{j}\big(\bar{h}_{ij}+\bar{\nabla}_{i}V_{j}+\bar{\nabla}_{j}V_{i}\big)-\bar{\nabla}_{i}\big(\bar{h}+2\bar{\nabla}_{j}V^{j}\big)\big]\Big]\,,\\0=&\frac{1}{2}e^{-(d-4)A(\rho)}\partial_{\rho}\Big[e^{dA(\rho)}\partial_{\rho}e^{-2A(\rho)}(\bar{h}+2\bar{\nabla}_{i}V^{i})\Big]\,,\label{eq:branegravitoneomsimplify}
    \end{split}
\end{equation}
where the upper indices are all lifted from lower indices using the inverse metric of the $d$-dimensional closed universe $\bar{g}^{ij}$. Near the Karch-Randall brane, the fields $\bar{h}_{ij}(x,\rho)$ and $V_{i}(x,\rho)$ and the gauge parameter $\epsilon_{i}(x,\rho)$ obey the following boundary conditions
\begin{equation}
    \partial_{\rho}e^{-2A(\rho)}\bar{h}_{ij}(x,\rho)_{\rho=\rho_{B}}=0\,,\quad\partial_{\rho}e^{-2A(\rho)}V_{i}(x,\rho)|_{\rho=\rho_{B}}=0\,,\quad\partial_{\rho}e^{-2A(\rho)}\epsilon_{i}(x,\rho)|_{\rho=\rho_{B}}=0\,.
\end{equation}
Therefore, we can expand both of them into Kaluza-Klein modes using the same set of Kaluza-Klein wavefunctions. The Kaluza-Klein wavefunctions are the eigen-solutions of the following differential equation and the same boundary condition as above
\begin{equation}
    -e^{-(d-4)A(\rho)}\partial_{\rho}e^{dA(\rho)}\partial_{\rho}e^{-2A(\rho)} \phi_{n}(\rho)=m_{n}^{2}\phi_{n}(\rho)\,,\quad\partial_{\rho}e^{-2A(\rho)}\phi_{n}(\rho)|_{\rho=\rho_{B}}=0\,.\label{eq:KKeomgraviton}
\end{equation}
We can work out the orthogonality condition of the wavefunctions $\phi_{n}(\rho)$ from the above differential equation and the boundary condition. The result is
\begin{equation}
    \int_{\rho_{B}}^{\infty} d\rho e^{(d-6)A(\rho)}\phi_{n}(\rho)\phi_{m}(\rho)=\delta_{nm}\,,\label{eq:normgraviton}
\end{equation}
for the unit normalized Kaluza-Klein wavefunctions. From Equ.~(\ref{eq:normgraviton}), we can already see that the zero mass eigenmode $\phi_{0}(\rho)=e^{2A(\rho)}=\cosh^{2}\rho$ is not normalizable. Thus, all the physical modes are massive. This is the essential physics revealed by the Karch-Randall braneworld \cite{Karch:2000ct} which has recently been refined in \cite{Geng:2025rov,Geng:2026wok}. Now we can work out the equations of motion for each Kaluza-Klein mode sector. For the $r$-th Kaluza-Klein mode sector, we have
\begin{equation}
    \begin{split}
       \mathcal{E}^{(d)}_{ijlm}\bar{h}^{(r)lm}-\frac{m_{r}^{2}}{2}(\bar{h}_{ij}^{(r)}-\bar{g}_{ij}\bar{h}^{(r)})-\frac{m_{r}^{2}}{2}(\bar{\nabla}_{i}V_{j}^{(r)}+\bar{\nabla}_{j}V_{i}^{(r)}-\bar{g}_{ij}2\bar{\nabla}_{i}V^{(r)i})&=0\,,\\
    \frac{m_{r}^{2}}{2}\Big[\bar{\nabla}^{j}\big(\bar{h}^{(r)}_{ij}+\bar{\nabla}_{i}V^{(r)}_{j}+\bar{\nabla}_{j}V^{(r)}_{i}\big)-\bar{\nabla}_{i}\big(\bar{h}^{(r)}+2\bar{\nabla}_{j}V^{(r)j}\big)\Big]&=0\,,\\
 m_{r}^{2}[\bar{h}^{(r)}+2\bar{\nabla}_{i}V^{(r)i}]&=0\,, 
 \end{split}
\end{equation}
where $m_{r}^{2}$ is the mass square of the $r$-th KK mode and the fields $\bar{h}^{(r)}$ and $V^{(r)i}$ are only functions of $x^{i}$, i.e. they are genuine $d$-dimensional fields. Defining the diffeomorphism invariant combination $\tilde{h}^{(r)}_{ij}$ as
\begin{equation}
    \tilde{h}^{(r)}_{ij}=\bar{h}^{(r)}_{ij}+\bar{\nabla}_{i}V^{(r)}_{j}+\bar{\nabla}_{j}V^{(r)}_{i}\,,
\end{equation}
we have
\begin{equation}
       \mathcal{E}^{(d)}_{ijlm}\tilde{h}^{(r)lm}-\frac{m_{r}^2}{2}(\tilde{h}_{ij}^{(r)}-\bar{g}_{ij}\tilde{h}^{(r)})=0\,,\quad\bar{\nabla}^{j}\tilde{h}^{(r)}_{ij}-\bar{\nabla}_{i}\tilde{h}^{(r)}=0\,,\quad\tilde{h}^{(r)}=0\,, 
\end{equation}
for which the last equation is redundant with the first equation as the linearized Einstein's equation is traceless. The first of the above equations is exactly the equation of motion of a $d$-dimensional massive graviton with a St\"{u}ckelberg mass term in the Fierz-Pauli form \cite{Hinterbichler:2011tt,deRham:2014zqa,deRham:2026pxl}. The second of the above equations is exactly the equation of motion of the St\"{u}ckelberg vector field $V^{(r)}_{i}$ from the St\"{u}ckelberg mass term of graviton. As a result, we can conclude that the brane localized gravity is purely a tower of spin-2 modes with S\"{u}ckelberg mass term. The lowest such mode can be identified as the localized graviton on the brane and its Kaluza-Klein wavefunction is indeed the only one that is localized on the Karch-Randall brane in the localization limit $\rho_{B}\rightarrow-\infty$. Finally, let's emphasize that the associated St\"{u}ckelberg/Goldstone vector fields are normalizable Kaluza-Klein modes of the extra-dimensional gravitational Wilson line \cite{Geng:2025rov}.

\subsubsection{The Emergence of the Observer and the Nontrivial Hilbert Space}
Having understood the induced gravitational theory on the brane, i.e. the closed gravitational universe, we are ready to see the emergence of the ``observer" and thus the nontrivial Hilbert space.

We follow the same logic as above, we will start by working out some properties of the bulk description and then map them to the intermediate description. For our purposes, let's use the bulk coordinate system Equ.~(\ref{eq:bulkmetric}). The brane is the hypersurface $\rho=\rho_{B}$, whose induced geometry is the closed universe Equ.~(\ref{eq:braneFRW}). We are interested in the Hamiltonian constraint. The Hamiltonian constraint is the $tt$-component of the bulk Einstein's equation. We would like to consider the situation with probe matter fields in the background geometry Equ.~(\ref{eq:bulkhbh}). The matter-sourced Hamiltonian constraint is given by
\begin{equation}
    \begin{split}
         0=&\sqrt{16\pi G_{N}}e^{4A(\rho)}\mathcal{H}_{\text{matter}}-\bar{\nabla}_{m}\bar{\nabla}_{n}\bar{h}^{mn}+\text{\ovA{$\Box$}} \bar{h}-\bar{g}^{tt}\bar{\nabla}_{t}\bar{\nabla}_{t}\bar{h}+2\bar{g}^{tt}\bar{\nabla}_{k}\bar{\nabla}_{t}\bar{h}_{t}^{k}-\bar{g}^{tt} \text{\ovA{$\Box$}}\bar{h}_{tt}-(d-1)(\bar{h}-2\bar{g}^{tt}\bar{h}_{tt})\,\\-&e^{-(d-4)A(\rho)}\partial_{\rho}\Big[e^{dA(\rho)}\partial_{\rho}e^{-2A(\rho)}(\bar{g}^{tt}\bar{h}_{tt}-\bar{h})\Big]+e^{-(d-4)A(\rho)}\partial_{\rho}\Big[e^{dA(\rho)}\partial_{\rho}e^{-2A(\rho)}(-2\bar{g}^{tt}\bar{\nabla}_{t}V_{t}+2\bar{\nabla}_{i}V^{i})\Big]\,,\label{eq:branegravitoneomsimplifysourcedfinal}
    \end{split}
\end{equation}
where we denote the matter Hamiltonian density associated with $t$ as $\mathcal{H}_{\text{matter}}$. This is a density in the $(d+1)$-dimensional bulk, and that of the brane localized matter is captured by delta-function terms of $\mathcal{H}_{\text{matter}}$. As explained in detail in \cite{Geng:2025rov,Geng:2025gqu}, we can set $\bar{h}_{t m}=0$ and the Hamiltonian constraint can be simplified as
\begin{equation}
\begin{split}
    -e^{-4A(\rho)}\Big[\frac{(d-2)\bar{h}}{\cos^2t}+\nabla_{a}\nabla_{b}\bar{h}^{ab}-\nabla^{2}\bar{h}-(d-2)\tan t\partial_{t}\bar{h}\Big]+\sqrt{16\pi G_{N}}\mathcal{H}_{\text{matter}}\\+e^{-dA(\rho)}\partial_{\rho}\Big[e^{dA(\rho)}\partial_{\rho}e^{-2A(\rho)}\Big(\bar{h}+2\nabla^{a}V_{a}-2(d-1)\tan t V^{t}\Big)\Big]=0\,.\label{eq:Hbrane}
    \end{split}
\end{equation}
The left-hand side of the Hamiltonian constraint Equ.~(\ref{eq:Hbrane}) contains three terms, and each of them has interesting features. As it was shown in Sec.~\ref{sec:noisometry}, if we consider the integrated Hamiltonian constraint for each KK mode, the first term gives $H_{\text{bulk}}[h^{(n)}]$; the second term is the local ``energy density" of the quantum field in a fixed background Equ.~(\ref{eq:bulkmetric}); the third term is the conjugate momentum of the $t$-component of the gravitational Wilson line $V_{t}(x,\rho)$,
\begin{equation}
     \pi_{V^{t}}(\vec{x},t,\rho)=-e^{-dA(\rho)}\partial_{\rho}\Big[e^{dA(\rho)}\partial_{\rho}e^{-2A(\rho)}\Big(\bar{h}+2\nabla^{a}V_{a}-2(d-1)\tan t V^{t}\Big)\Big]\,,\label{eq:Pi1}
\end{equation}
which by canonical quantization gives us the equal-time commutator
\begin{equation}
  [V^{t}(\vec{y},t,\rho),\pi_{V^{t}}(\vec{x},t,\rho')]=i\frac{1}{e^{dA(\rho)}\sqrt{-\bar{g}}}\delta(\rho-\rho')\delta^{d-1}(\vec{y}-\vec{x})\,.\label{eq:Pi2}
\end{equation}

As we discussed in Sec.~\ref{sec:noisometry}, the effects associated with the first term in Equ.~(\ref{eq:Hbrane}) is not important for our purposes, so we will drop this term hereafter. Thus, if one considers the integrated Hamiltonian constraint, with the integration performed along the bulk Cauchy slice, one would have
\begin{equation}
    \sqrt{16\pi G_{N}}H_{\text{matter}}-H_{B}=0\,,\label{eq:bulkHconstraint}
\end{equation}
where we have defined
\begin{equation}
    \begin{split}
        H_{B}\equiv \Pi_{V^{t}}&=\int e^{dA(\rho)}\sqrt{-\bar{g}} d\rho d^{d-1}\vec{x}\pi_{V^{t}}(x,\rho)\,,\\&=-\lim_{\rho_{c}\rightarrow\infty}\int \sqrt{-\bar{g}}d^{d-1}\vec{x}e^{dA(\rho_{c})}\partial_{\rho}e^{-2A(\rho_{c})}\Big(\bar{h}+2\nabla_{a}V^{a}-2(d-1)\tan t V^{t}\Big)(x,\rho_{c})\,,\label{eq:redef}
    \end{split}
\end{equation}
where $x=(t,\vec{x})$. In fact, this result Equ.~(\ref{eq:bulkHconstraint}) is nothing but the gravitational Gauss's law in the bulk AdS$_{d+1}$, which states that energy sourced by matter fields in the bulk, including on the brane, can be detected on the asymptotic boundary where gravity turns off. In the quantum case, the equation Equ.~(\ref{eq:bulkHconstraint}) should be understood as an operator equation, which constrains operators and states. For example, a matter sector operator $\hat{\phi}(x)$\footnote{We should be a bit more careful here. The argument $x$ is actually $x^{\mu}=(x^{i},\rho)$ which is a simplified notation. For the sake of convenience, we will call the argument of all operators as $x$, no matter for a bulk operator which depends on $x^\mu$, for a brane-localized operator which only depends on $x^{i}$ or for the operators in the intermediate description that we will discuss later-on. We hope the meaning will be clear given the context.} is not an observable in the gravitational theory by itself, because it doesn't obey the constraint
\begin{equation}
    [\sqrt{16\pi G_{N}}\hat{H}_{\text{matter}}-\hat{H}_{B},\hat{O}^{\text{Phys}}(x)]=0\,,\label{eq:bulkHconstraintO}
\end{equation}
where now we use $\hat{}$ to denote operators which shouldn't be confused with the $\hat{}$ we used for the metric on $\Sigma_{d-1}$. However, solutions of of Equ.~(\ref{eq:bulkHconstraintO}) can be constructed by properly dressing the operator $\hat{\phi}(x)$. Using Equ.~(\ref{eq:Pi1}), Equ.~(\ref{eq:Pi2}) and Equ.~(\ref{eq:redef}), it is easy to see that solutions of the above constraint are of the form
\begin{equation}
    \hat{O}^{\text{Phys}}(x)=\hat{\phi}(x+\sqrt{16\pi G_{N}}\hat{V}(x))\,,
\end{equation}
i.e. the argument is the invariant combination $x^{\mu}+\sqrt{16\pi G_{N}}\hat{V}^{\mu}(x)$ and the operator is dressed using the gravitational Wilson line $V^{\mu}(x)$ from the bulk to the asymptotic boundary. This is analogous to the Wilson line dressed operators in gauge theory. By the AdS/CFT correspondence, $\hat{H}_{B}$ is identified as the dual CFT Hamiltonian, and the above constraint equation states that an energetic excitation in the bulk is also an energetic excitation for the CFT.\footnote{Here we emphasize that there are also energy excitations in the bulk which are not (visibly) detectable on the boundary in this sense. These excitations are produced by operators of the form Equ.~(\ref{eq:Opert}). They obey the constraint due to the first term in Equ.~(\ref{eq:Hbrane}). As we have discussed, these operators are not important for our purposes, so we ignore them hereafter.} This standard understanding in holography can be rephrased using the idea of ``observer" in the recent literature \cite{Witten:2023qsv,Witten:2023xze,Chandrasekaran:2022cip,Harlow:2025pvj,Geng:2025gqu} as: 
\begin{displayquote}
\textit{The asymptotic boundary of the bulk AdS, i.e. the dual CFT, is an observer and the physics inside the gravitational bulk is nontrivial as it is defined relative to this observer.} 
\end{displayquote}
In other words, the existence of the ``observer" ensures that the Hilbert space associated with the gravitational bulk is nontrivial. This manifests itself if one notices that if for some reason the ``observer" term $\hat{H}_{B}$ in Equ.~(\ref{eq:bulkHconstraintO}) disappears, the only solution for that operator constraint would be $c$-numbers. This heuristically indicates that the underlying Hilbert space becomes trivially one-dimensional.

Now we can map the above results in the bulk description to the intermediate description. From the intermediate description perspective, a probe operator in the closed gravitational universe can be thought of as being coupled only to a few Kaluza-Klein gravitons. For simplicity, we focus on the case that an intermediate description operator $\hat{O}(x)$ only couples to the $n$-th Kaluza-Klein graviton. As a result, this intermediate description operator has to obey the constraint Equ.~(\ref{eq:Hbrane}) projected to the $n$-th Kaluza-Klein mode. The integrated version of this intermediate description constraint is 
\begin{equation}
    [\sqrt{16\pi G_{N}}\hat{H}_{\text{matter}}^{(n)}-m_{n}^{2}\int \sqrt{-\bar{g}}d^{d-1}\vec{x}\Big(h^{(n)}+2\nabla_{i}V^{(n)i}-2(d-1)\tan t V^{(n)t}\Big),\hat{O}^{\text{Phys}}(x)]=0\,,\label{eq:constraintbrane}
\end{equation}
where we defined
\begin{equation}
    \hat{H}_{\text{matter}}^{(n)}\equiv\int \sqrt{-\bar{g}}d^{d-1}\vec{x}d\rho e^{(d-2)A(\rho)}\phi_{n}(\rho)\mathcal{H}_{\text{matter}}\,,
\end{equation}
which is the Hamiltonian that evolves the operator $\hat{O}(x)$. Moreover, we also notice that the conjugate momentum for the $\xi$-component of the $n$-th St\"{u}ckelberg vector field $V^{(n)i}(x)$ is
\begin{equation}
    \pi_{V^{t}}^{(n)}(\vec{x},t)=m_{n}^{2}\Big(h^{(n)}+2\nabla^{a}V^{(n)}_{a}-2(d-1)\tan t V^{(n)t}\Big)(\vec{x},t)\,.\label{eq:pin}
\end{equation}
Hence, Equ.~(\ref{eq:constraintbrane}) can be written as
\begin{equation}
     [\sqrt{16\pi G_{N}}\hat{H}_{\text{matter}}^{(n)}-\Pi^{(n)}_{V^{t}},\hat{O}^{\text{Phys}}(x)]=0\,.\label{eq:braneconstraintsimpl}
\end{equation}
Nontrivial solutions of this equation are of the form
\begin{equation}
    \hat{O}^{\text{Phys}}(x)=\hat{O}(x+\sqrt{16\pi G_{N}}\hat{V}^{(n)}(x))\,,\label{eq:KRislando}
\end{equation}
which are genuine operators in the closed gravitational universe. From the perspective of the closed gravitational universe, these operators are locally dressed by the Goldstone vector field $V^{(n)i}(x)$. This suggests the following:
\begin{displayquote}
\textit{We should interpret the Goldstone vector boson as an ``observer" field which could sense locations in the spacetime and the physical observables are operators dressed to this ``observer". }
\end{displayquote}
These dressed operators are nontrivial operators and so the underlying Hilbert space they act on cannot be trivially one-dimensional. This is thus an explicit realization of the proposals that nontrivial Hilbert space in a closed gravitational universe could be induced by an ``observer" in the recent literature \cite{Witten:2023qsv,Witten:2023xze,Chandrasekaran:2022cip,Harlow:2025pvj,Geng:2025gqu}. We will discuss more about this point in Sec.~\ref{sec:comments}. 

Moreover, we also have
\begin{equation}
\begin{split}
[\hat{H}_{\text{CFT}},\hat{O}^{\text{Phys}}(x)]&=[\frac{1}{\sqrt{16\pi G_{N}}}\hat{H}_{B},\hat{O}(x+\sqrt{16\pi G_{N}}\hat{V}^{(n)}(x))]\,,\\&=-i\partial_{t}\hat{O}^{\text{Phys}}(x)\int_{\rho_{B}}^{\infty}d\rho e^{(d-6)A(\rho)} \phi_{n}(\rho)\neq0\,,\label{eq:nonlocal}
\end{split}
\end{equation}
where we ignored terms of order $\mathcal{O}(G_{N})$ in the final result. In this result, the $\rho$ integral is finite because normalizable modes decay as $e^{-(d-2)A(\rho)}$ as $\rho\rightarrow\infty$ \cite{Lashkari:2013koa}. This result is clearly consistent with the boundary description, as eventually the fundamental Hilbert space is the dual CFT Hilbert space so any nontrivial operator would excite the CFT Hamiltonian. Nevertheless, we should note that the above effect manifests itself already at the perturbative level in $G_{N}$. Thus, one would expect that in the intermediate description, where we have a closed gravitational universe in which some matter fields are entangled with an external non-gravitational system, the result Equ.~(\ref{eq:nonlocal}) suggests that an observer in that external non-gravitational system could detect excitations created inside the closed gravitational universe by the operator $\hat{O}^{\text{Phys}}(x)$ by simply measuring the energy. Thus, this is a genuinely non-local effect, which manifests in the intermediate description at the level of low-energy gravitational effective field theory. As it was understood in \cite{Geng:2025byh,Geng:2025gqu}, this effect is due to the interplay between dynamical gravity and the quantum entanglement of the matter fields in the intermediate description, and it shares the same origin as the graviton mass and entanglement islands. More importantly, the nonzero commutator in Equ.~(\ref{eq:nonlocal}) is not suppressed by any power of the gravitational coupling $G_{N}$. Thus, this is the sharp nonperturbative signature that distinguishes the operators of the form Equ.~(\ref{eq:KRislando}) from the operators like Equ.~(\ref{eq:Opert}). More precisely, operators of the form Equ.~(\ref{eq:KRislando}) are dressed to the quantum entanglement between the closed gravitational universe and the external non-gravitational system, which is a nonperturbative feature of the quantum state.

\subsection{In the Intermediate Description}\label{sec:intermediate}
Let's consider the following model of the intermediate description that in the $d$-dimensional closed gravitational universe with metric
\begin{equation}
    ds^2=-dt^2+\cos^2 t d\Sigma_{d-1}^2\,, \label{eq:Lorenzian}
\end{equation}
we have Einstein's gravity weakly coupled with a massive free scalar field $\phi_{1}(x)$ with a mass squared obeying $-\frac{(d-1)^2}{4}<m^2<1-\frac{(d-1)^2}{4}$. There are two maximally symmetric (pure) states of this scalar field, which are defined by the time-ordered two-point functions
\begin{equation}
\begin{split}
\langle \mathbf{T}\phi_{1}(x_{1})\phi_{1}(x_{2})\rangle&=\sum_{\gamma\in\Gamma}\Bigg( a_{1}2^{-\Delta_{1}}\frac{\Gamma[\Delta_{1}]}{\pi^{\frac{d-1}{2}}(2\Delta_{1}-d+1)\Gamma[\Delta_{1}-\frac{d-1}{2}]}\frac{_{2}F_{1}[\frac{\Delta_{1}}{2},\frac{\Delta_{1}+1}{2},\Delta_{1}-\frac{d-1}{2}+1,\frac{1}{Z_{\gamma}^{2}}]}{(Z_\gamma)^{\Delta_{1}}}\\&+a_{2}2^{-\Delta_{2}}\frac{\Gamma[\Delta_{2}]}{\pi^{\frac{d-1}{2}}(2\Delta_{2}-d+1)\Gamma[\Delta_{2}-\frac{d-1}{2}]}\frac{_{2}F_{1}[\frac{\Delta_{2}}{2},\frac{\Delta_{2}+1}{2},\Delta_{2}-\frac{d-1}{2}+1,\frac{1}{Z_{\gamma}^{2}}]}{(Z_{\gamma})^{\Delta_{2}}}\Bigg)\,,
\end{split}
\end{equation}
where the two states correspond respectively to $(a_{1},a_{2})=(1,0)$ and $(a_{1},a_{2})=(0,1)$, $\Delta_{1}=(d-1)-\Delta_{2}=\frac{d-1}{2}+\sqrt{\frac{(d-1)^2}{4}+m^2}$ and $Z_{\gamma}$ is the invariant-distance between the point $x_{1}$ and the image of the point $x_{2}$ under $\gamma$.

In our case, we are interested in the entangled state between two such closed universes with one gravitational and the other non-gravitational. Let's model the matter field that is entangled with $\phi_{1}(x)$ in the non-gravitational closed universe as a free massive scalar field $\phi_{2}(x)$ with equal mass as $\phi_{1}(x)$.\footnote{More precisely, $\phi_{1}$ and $\phi_{2}$ have the same value of the product of the mass and the respective AdS length scale. For convenience, we set all AdS length scales to one.} This entangled state is prepared by the Euclidean path integral on a manifold which is obtained by gluing two manifolds with the metric
\begin{equation}
    ds_{E}^2=d\tau^2+\cosh^2\tau d\Sigma_{d-1}^2\,,\label{eq:Euclidean1}
\end{equation}
at $\tau\rightarrow-\infty$. The state is defined on the union of the $\tau=0$ slices of the two constituent manifolds. This closely models the intermediate description of the braneworld setup we considered. For clarity, we only consider conformally coupled scalars $\phi_{1}(x)$ and $\phi_{2}(x)$ with mass squared $m^2=-\frac{(d-1)^2-1}{4}$.\footnote{We note that in fact conformally coupled scalar field in compact hyperbolic geometries has a normalizable imaginary frequency mode \cite{Buchel:2004rr}. This drives the state prepared by the Euclidean path integral on this type of geometries non-normalizable. We ignore this issue in this paper as we only take the conformally coupled scalar field as a particular example to demonstrate our calculation. The same calculation will go through for scalar fields with other masses.} For a conformally coupled scalar, the metric Equ.~(\ref{eq:Euclidean1}) of the state-preparation manifold can be replaced by the Weyl-equivalent metric
\begin{equation}
    ds=d\tilde{\eta}^2+d\Sigma_{d-1}^{2}\,,\label{eq:Euclidean2}
\end{equation}
where $\tilde{\eta}\in[-\pi,0]$ and the entangled state is supported on the union of the slices $\tilde{\eta}=0,-\pi$. The two scalar fields $\phi_{1}(x)$ and $\phi_{2}(x)$ can be identified as one conformal scalar field on Equ.~(\ref{eq:Euclidean2}) where $\phi_{1}(x)$ describes the part $\tilde{\eta}\in[-\pi,-\frac{\pi}{2})$ and $\phi_{2}(x)$ describes the part $\tilde{\eta}\in(-\frac{\pi}{2},0]$ (as depicted in Fig.~\ref{pic:intermediatepathintegral} where the black dashed circle is $\tilde{\eta}=-\frac{\pi}{2}$). We are interested in the state prepared by this Euclidean path integral, without any nontrivial defect inserted on the $\tilde{\eta}=-\frac{\pi}{2}$-slice.  Translating back to the original manifold Equ.~(\ref{eq:Euclidean1}) or Equ.~(\ref{eq:Lorenzian}), this means that we have the following correlators
\begin{equation}
\begin{split}
\langle \mathbf{T}\phi_{1}(x_{1})\phi_{1}(x_{2})\rangle&=\sum_{\gamma\in\Gamma}\frac{1}{2}\Bigg( 2^{-\Delta_{1}}\frac{\Gamma[\Delta_{1}]}{\pi^{\frac{d-1}{2}}(2\Delta_{1}-d+1)\Gamma[\Delta_{1}-\frac{d-1}{2}]}\frac{_{2}F_{1}[\frac{\Delta_{1}}{2},\frac{\Delta_{1}+1}{2},\Delta_{1}-\frac{d-1}{2}+1,\frac{1}{Z_{\gamma}^{2}}]}{(Z_\gamma)^{\Delta_{1}}}\\&+2^{-\Delta_{2}}\frac{\Gamma[\Delta_{2}]}{\pi^{\frac{d-1}{2}}(2\Delta_{2}-d+1)\Gamma[\Delta_{2}-\frac{d-1}{2}]}\frac{_{2}F_{1}[\frac{\Delta_{2}}{2},\frac{\Delta_{2}+1}{2},\Delta_{2}-\frac{d-1}{2}+1,\frac{1}{Z_{\gamma}^{2}}]}{(Z_{\gamma})^{\Delta_{2}}}\Bigg)\,,\label{eq:phi1phi1}
    \end{split}
\end{equation}

\begin{equation}
\begin{split}
    \langle \mathbf{T}\phi_{2}(x_{1})\phi_{2}(x_{2})\rangle&=\sum_{\gamma\in\Gamma}\frac{1}{2}\Bigg( 2^{-\Delta_{1}}\frac{\Gamma[\Delta_{1}]}{\pi^{\frac{d-1}{2}}(2\Delta_{1}-d+1)\Gamma[\Delta_{1}-\frac{d-1}{2}]}\frac{_{2}F_{1}[\frac{\Delta_{1}}{2},\frac{\Delta_{1}+1}{2},\Delta_{1}-\frac{d-1}{2}+1,\frac{1}{Z_{\gamma}^{2}}]}{(Z_\gamma)^{\Delta_{1}}}\\&+2^{-\Delta_{2}}\frac{\Gamma[\Delta_{2}]}{\pi^{\frac{d-1}{2}}(2\Delta_{2}-d+1)\Gamma[\Delta_{2}-\frac{d-1}{2}]}\frac{_{2}F_{1}[\frac{\Delta_{2}}{2},\frac{\Delta_{2}+1}{2},\Delta_{2}-\frac{d-1}{2}+1,\frac{1}{Z_{\gamma}^{2}}]}{(Z_{\gamma})^{\Delta_{2}}}\Bigg)\,,
    \end{split}
\end{equation}
\begin{equation}
\begin{split}
    \langle \mathbf{T}\phi_{1}(x_{1})\phi_{2}(x_{2})\rangle&=\sum_{\gamma\in\Gamma}\frac{1}{2}\Bigg( 2^{-\Delta_{1}}\frac{\Gamma[\Delta_{1}]}{\pi^{\frac{d-1}{2}}(2\Delta_{1}-d+1)\Gamma[\Delta_{1}-\frac{d-1}{2}]}\frac{_{2}F_{1}[\frac{\Delta_{1}}{2},\frac{\Delta_{1}+1}{2},\Delta_{1}-\frac{d-1}{2}+1,\frac{1}{Z_{\gamma}^{2}}]}{(Z_\gamma)^{\Delta_{1}}}\\&-2^{-\Delta_{2}}\frac{\Gamma[\Delta_{2}]}{\pi^{\frac{d-1}{2}}(2\Delta_{2}-d+1)\Gamma[\Delta_{2}-\frac{d-1}{2}]}\frac{_{2}F_{1}[\frac{\Delta_{2}}{2},\frac{\Delta_{2}+1}{2},\Delta_{2}-\frac{d-1}{2}+1,\frac{1}{Z_{\gamma}^{2}}]}{(Z_{\gamma})^{\Delta_{2}}}\Bigg)\,,
    \end{split}
\end{equation}
where the last correlator indicates the nontrivial entanglement between the two closed universes and the invariant-distance $Z$ is computed as if $\phi_{1}$ and $\phi_{2}$ are living in the same closed universe. We are interested in the gravitational part of the Euclidean path integral that computes the norm of the above entangled state. This path integral can be written as
\begin{equation}
\begin{split}
Z_{H}&=\int D[\phi_{1}]D[h_{\mu\nu}]e^{-S_{\text{matter}}[\phi_1;\bar{g}]-S[h;\bar{g}]-S_{\text{int}}[\phi_1,h;\bar{g}]}\,,\\S_{\text{int}}[\phi_1,h;\bar{g}]&=\frac{\sqrt{16\pi G_{d}}}{2}\int d^{d}x\sqrt{\bar{g}(x)}h_{ij}(x)T^{ij}_{\text{matter}}(x)\,,\label{eq:Z}
\end{split}
\end{equation}
where $\bar{g}_{ij}$ denotes the background geometry Equ.~(\ref{eq:Euclidean1}) with $\tau\in(-\infty,\infty)$, $G_{d}$ is the Newton's constant in the closed gravitational universe, $\sqrt{16\pi G_{d}}h_{ij}$ denotes the fluctuation of this background metric, i.e. $h_{ij}$ is the graviton field, and $S[h;\bar{g}]$ denotes the kinetic term of the graviton. We are only interested in leading order gravitational effects, and to this leading order in $G_{d}$, the diffeomorphism transforms act as
\begin{equation}
\begin{split}
&x^{i}\rightarrow x^{i}, \quad \bar{g}_{ij}(x)\rightarrow \bar{g}^{\prime}_{ij}(x)=\frac{\partial x'^{l}}{\partial x^{i}}\frac{\partial x'^{m}}{\partial x^{j}}\bar{g}_{lm}(x')\,,\quad\phi_{1}(x)\rightarrow\phi'_{1}(x)=\phi_{1}(x')\,,\\&h_{ij}(x)\rightarrow h_{ij}'(x)=\frac{\partial x'^{l}}{\partial x^{i}}\frac{\partial x'^{m}}{\partial x^{j}}h_{lm}(x')+\bar{\nabla}_{i}\epsilon_{j}(x)+\bar{\nabla}_{j}\epsilon_{i}(x)\,,\quad \text{where } x'^{i}=x^{i}+\sqrt{16\pi G_{d}}\epsilon^{i}(x)\,.\label{eq:diffeotr}
\end{split}
\end{equation}
The full action $S_{\text{full}}[\phi_{1},h;\bar{g}]=S_{\text{matter}}[\phi_{1};\bar{g}]+S[h;\bar{g}]+S_{\text{int}}[\phi_{1},h;\bar{g}]$ in Equ.~(\ref{eq:Z}) transforms as
\begin{equation}
    S_{\text{full}}[\phi_{1},h;\bar{g}]\rightarrow  S_{\text{full}}[\phi_{1},h;\bar{g}]+\sqrt{16\pi G_{d}}\int d^{d}x\sqrt{\bar{g}(x)}\bar{\nabla}_{i}\epsilon_{j}(x) T^{ij}_{\text{matter}}(x)\,,\label{eq:Sfulltransform}
\end{equation}
where we have used the fact that $T^{ij}_{\text{matter}}(x)$ is a symmetric tensor. In our case of the closed gravitational universe, there is a boundary in the Euclidean signature as $\tau\rightarrow-\infty$\footnote{And also $\tau=\infty$.} and this is the place where the entanglement between the matter field $\phi_{1}$ in this gravitational universe and the matter field $\phi_2$ in the other non-gravitational closed universe is generated. This manifests itself in the Weyl-equivalent manifold Equ.~(\ref{eq:Euclidean2}) where this boundary is transparent to the matter fields. Thus, as a result, the Euclidean stress-energy flux for the gravitational part is not zero near this boundary and we have
\begin{equation}
\begin{split}
    &\int d^{d}x\sqrt{\bar{g}(x)}\bar{\nabla}_{i}\epsilon_{j}(x) T^{ij}_{\text{matter}}(x)\,,\\&=\int d^{d}x\sqrt{\bar{g}(x)}\bar{\nabla}_{i}\Big(\epsilon_{j}(x) T^{ij}_{\text{matter}}(x)\Big)-\int d^{d}x\sqrt{\bar{g}(x)}\epsilon_{j}(x) \bar{\nabla}_{i}T^{ij}_{\text{matter}}(x)\,,\\&=\int d^{d-1}x\sqrt{\bar{g}(x)}\epsilon_{j}(x) T^{i\perp}_{\text{matter}}(x)\neq0\,,\label{eq:nonvanishing}
    \end{split}
\end{equation}
where the second term in the second line is zero due to the local conservation of the stress-energy tensor inside the closed gravitational universe. The non-vanishing result in Equ.~(\ref{eq:nonvanishing}) together with Equ.~(\ref{eq:Sfulltransform}) indicates that the full action is not manifestly diffeomorphism invariant. To have a manifestly covariant, i.e. diffeomorphism invariant, description, we can introduce a compensating auxiliary vector field $V^{i}(x)$ which transforms under the diffeomorphism Equ.~(\ref{eq:diffeotr}) as
\begin{equation}
    V_{i}(x)\rightarrow V'_{i}(x)=\frac{\partial x^{l\prime}}{\partial x^{i}} V_{l}(x')-\epsilon_{i}(x)\,.\label{eq:diffV}
\end{equation}
We integrate the auxiliary vector into the path integral Equ.~(\ref{eq:Z}) in the following way
\begin{equation}
    Z_{\text{full}}=\int D[V^{i}]D[\phi_{1}]D[h_{ij}]e^{-S_{\text{matter}}[\phi_{1};\bar{g}]-S[h;\bar{g}]-S_{\text{int}}[\phi_{1},h;\bar{g}]-\sqrt{16\pi G_{d}}\int d^{d}x\sqrt{\bar{g}(x)}\bar{\nabla}_{i}V_{j}(x)T^{ij}_{\text{matter}}(x)}\,.\label{eq:Zfull}
\end{equation}
As a result, using Equ.~(\ref{eq:Sfulltransform}) together with Equ.~(\ref{eq:diffV}), it is easy to see that this full partition function is manifestly invariant under the diffeomorphism transform to leading order in $G_{d}$. Furthermore, the original partition function Equ.~(\ref{eq:Z}) is now a gauge-fixed version of Equ.~(\ref{eq:Zfull}) under the unitary gauge $V_{i}(x)=0$. As is emphasized in \cite{Geng:2025rov}, this observation is parallel to the standard story in the Higgs mechanism, where the unitary gauge fixes the Goldstone boson to zero such that the Goldstone boson disappears in the path integral, which could be interpreted as being eaten by the massive gauge boson \cite{peskin1995iqf}. This suggests that the gravitational theory in the closed gravitational universe should be in the Higgs phase such that the graviton is massive. This is consistent with what we have learned from the the bulk description in Sec.~\ref{sec:bulk}. To prove this, we have to show that there is an induced St\"{u}ckelberg mass term for the graviton $h^{ij}(x)$ with $V^{i}(x)$ as the associated St\"{u}ckelberg/Goldstone vector field. This St\"{u}kelberg mass term should be a result of quantum loop effect from the field $\phi_{1}(x)$. To see the emergence of such a mass term, we will integrate out $\phi_{1}(x)$ to get an effective action for the graviton $h^{ij}(x)$ and the vector field $V^{i}(x)$. As a result of the manifest diffeomorphism invariance, this effective action should also be invariant under the transform Equ.~(\ref{eq:diffeotr}) together with Equ.~(\ref{eq:diffV}). Hence, this effective action could only be a functional of the invariant combination
\begin{equation}
    h_{ij}(x)+\bar{\nabla}_{i}V_{j}(x)+\bar{\nabla}_{j}V_{i}(x)\,.\label{eq:invariant}
\end{equation}
For our purposes, we only have to focus on the quadratic terms of this invariant combination in the resulting effective action in the low-energy regime. Following the strategy in \cite{Geng:2023ynk,Geng:2023zhq,Geng:2025rov}, we will first send the graviton field to zero, only consider the interaction between $\phi_{1}(x)$ and the auxiliary vector field $V^{\mu}(x)$, and replace $\bar{\nabla}_{i}V_{j}(x)+\bar{\nabla}_{j}V_{i}(x)$ in the resulting effective action with the invariant combination Equ.~(\ref{eq:invariant}). Thus, first we have the induced action for the auxiliary vector field
\begin{equation}
\begin{split}
    \delta S_{2}[V]=-\frac{1}{2}(\sqrt{16\pi G_{d}})^{2}\int d^{d}xd^{d}y\sqrt{\bar{g}(x)}\sqrt{\bar{g}(y)}\bar{\nabla}_{i}V_{j}(x)\bar{\nabla}_{l}V_{m}(y)\langle \mathbf{T} T^{ij}(x) T^{lm}(y)\rangle\,,\label{eq:Vbare}
\end{split}
\end{equation}
where $\mathbf{T}$ denotes the time-ordering in Euclidean time $\tau$ and  $T^{ij}(x)$ is the stress-energy tensor of $\phi_{1}(x)$. The low-energy regime can be isolated by taking the large invariant-distance limit of the two-point function 
\begin{equation}
    \Pi^{ij,lm}(x,y)=\langle \mathbf{T} T^{ij}(x) T^{lm}(y)\rangle\,.\label{eq:TT2pt}
\end{equation}
We note that as opposed to the computations in the open AdS/CFT \cite{Porrati:2001gx,Aharony:2006hz,Geng:2025rov,Baumann:2026lnt}, here the large invariant-distance limit is slightly subtler, as the spatial manifold $\Sigma_{d-1}$ is compact. We would consider the limit $\tau_{x},\tau_{y}\rightarrow-\infty$, i.e. near the boundary, and then extend the result into the bulk by general covariance. In this near boundary regime, we are automatically in the large invariant-distance limit, as the invariant-distance for two different points $x$ and $y$ are blowing up as $\tau_{x},\tau_{y}\rightarrow-\infty$ with the spatial distance on $\Sigma_{d-1}$ fixed in taking this limit. Another subtlety is that in our case the two-point correlator Equ.~(\ref{eq:phi1phi1}) contains contributions from many images of the discrete group $\Gamma$. These contributions might complicate the result of Equ.~(\ref{eq:TT2pt}). However, in fact for our purposes of deriving a local effective action of $V^{i}(x)$, we can ignored all those contributions in Equ.~(\ref{eq:phi1phi1}) other than $\gamma=\mathbf{1}$. The reason for this will become clear soon. It is nicely computed in \cite{Aharony:2006hz} that for the contribution from $\gamma=\mathbf{1}$ the large distance limit of the stress-energy tensor two point function is 
\begin{equation}
    \begin{split}
        16\pi G_{d}\Pi^{ij,lm}(x,y)\supset 4M^{2}\bar{\nabla}^{(i}D^{j),(l}(x,y)\bar{\nabla}^{m)}\,,\label{eq:TT}
        \end{split}
\end{equation}
where $D^{j,l}(x,y)$ is the propagator of a massive vector field with mass squared $m_{V}^2=2(d-1)$ which satisfies
\begin{equation}
(\bar{\nabla}^{2}-d+1)D^{i,j}(x,y)=\frac{\bar{g}^{ij}}{\sqrt{\bar{g}}}\delta^{d}(x-y)\,.\label{eq:Green}
\end{equation}
The value of the graviton mass $M^{2}$ is obtained in \cite{Aharony:2006hz} as
\begin{equation}
M^{2}=-G_{d}\frac{2^{5-d}\pi^{\frac{4-d}{2}}}{(d+1)\Gamma(\frac{d+2}{2})}\frac{a_{1}a_{2}\Delta_{1}\Delta_{2}\Gamma[\Delta_{1}]\Gamma[\Delta_{2}]}{\Gamma[\Delta_{1}-\frac{d-1}{2}]\Gamma[\Delta_{2}-\frac{d-1}{2}]}\,,\label{eq:mass}
\end{equation}
where we have $a_{1}=a_{2}=\frac{1}{2}$ and $\Delta_{1}=(d-1)-\Delta_{2}=\frac{d}{2}$. We note that the propagator $D^{\nu,\sigma}(x,y)$ is divergenceless as a massive vector field is automatically divergenceless on-shell. Using the above results and integrating by parts, we have
\begin{equation}
    \begin{split}
    \delta S_{2}[V]&=\frac{1}{2}(\sqrt{16\pi G_{d}})^{2}\int d^{d}xd^{d}y\sqrt{\bar{g}(x)}\sqrt{\bar{g}(y)}V_{j}(x)\bar{\nabla}_{l}V_{m}(y)\nabla_{i}\langle \mathbf{T} T^{ij}(x) T^{lm}(y)\rangle\,,\\&=-\frac{1}{2}2M^{2}\int d^{d}x\sqrt{\bar{g}(x)}V_{j}(x)\bar{g}^{i(m}(x)\bar{\nabla}^{l)}\bar{\nabla}_{l}V_{m}(x)\,,\\&=-\frac{M^{2}}{2}\int d^{d}x\sqrt{\bar{g}(x)}\Big[V_{j}(x)\bar{\nabla}^{i}\bar{\nabla}_{i}V^{j}(x)+V_{j}(x)\bar{\nabla}^{i}\bar{\nabla}^{j}V_{i}(x)\Big]\,,\\&=\frac{M^{2}}{2}\int d^{d}x\sqrt{\bar{g}(x)}\Big[\bar{\nabla}^{i}V^{j}(x)\bar{\nabla}_{i}V_{j}(x)+\bar{\nabla}^{i}V^{j}(x)\bar{\nabla}_{j}V_{i}(x)\Big]\,,\\&=\frac{M^{2}}{4}\int d^{d}x\sqrt{\bar{g}(x)}\Big(\bar{\nabla}^{i}V^{j}(x)+\bar{\nabla}^{j}V^{i}(x)\Big)\Big(\bar{\nabla}_{i}V_{j}(x)+\bar{\nabla}_{j}V_{i}(x)\Big)\,.\label{eq:Valmost}
\end{split}
\end{equation}
Here we can see why we can ignore the contributions from the other images of the discrete group $\Gamma$ to the two-point function Equ.~(\ref{eq:TT2pt}). This is because to work out the effective action we are only interested in the nonzero delta-function terms of
\begin{equation}
    \nabla_{i}\langle \mathbf{T} T^{ij}(x) T^{lm}(y)\rangle\,,\label{eq:DTT}
\end{equation}
and the contributions from the the images associated with other group elements of $\Gamma$ to the two-point function Equ.~(\ref{eq:TT2pt}) would contribute trivially to Equ.~(\ref{eq:DTT}).\footnote{This is the standard mirror charge story as in electrodynamics, where the mirror charges are introduced such that the Green's function obeys the correct boundary equation but the $\delta$-function source of the Green's equation is not changed by these extra terms.} Furthermore, due to the fact that the propagator $D^{i,j}(x,y)$ is divergenceless, the two-point function Equ.~(\ref{eq:TT}) is traceless in both $ij$ and $lm$. Therefore, the divergence of the vector field $\nabla_{i}V^{i}(x)$ should decouple. As a result, in the correct effective action $\delta S_{2}[V]$, the divergence part of the vector field should decouple. In other words, the propagator of the vector field $V^{i}(x)$ has to be divergenceless. Thus, the correct off-shell effective action should be
\begin{equation}
    \delta S_{2}[V]=\frac{M^{2}}{4}\int d^{d}x\sqrt{\bar{g}(x)}\Big[\Big(\bar\nabla^{i}V^{j}(x)+\bar\nabla^{j}V^{i}(x)\Big)\Big(\bar\nabla_{i}V_{j}(x)+\bar\nabla_{j}V_{i}(x)\Big)-4\bar\nabla^{i}V_{i}(x)\bar\nabla^{j}V_{j}(x)\Big]\,.\label{eq:Vfinal}
\end{equation} 
As a consistent check, one can work out the equation of motion of the vector field from Equ.~(\ref{eq:Vfinal})
\begin{equation}
    \bar\nabla^{2}V^{j}(x)+\bar\nabla_{i}\bar\nabla^{j}V^{i}(x)-2\bar\nabla^{j}\nabla_{i}V^{i}(x)=0\,,
\end{equation}
which can be rewritten as
\begin{equation}
    \bar\nabla_{i}(\bar\nabla^{i}V^{j}-\bar\nabla^{j}V^{i})(x)-2(d-1) V^{j}(x)=0\,,
\end{equation}
where we used $R_{ij}=-(d-1)\bar{g}_{ij}$ for AdS$_{d}$. Thus, we indeed obtain the equation of motion of a massive vector field and it is easy to see that $V^{i}(x)$ is divergenceless using this equation of motion. Finally, we are ready to restore the $h_{ij}(x)$ dependence of the effective action by the replacement by the invariant combination in the same form as Equ.~(\ref{eq:invariant}). This gives
\begin{equation}
    S_{\text{eff}}^{(2)}[h,V]=\frac{M^{2}}{4}\int d^{d}x\sqrt{\bar{g}(x)}\Big(\tilde{h}_{ij}(x)\tilde{h}^{ij}(x)-\tilde{h}^{2}(x)\Big)\,,\label{eq:stuckelberg}
\end{equation}
where $\tilde{h}_{ij}(x)$ denotes the invariant combination Equ.~(\ref{eq:invariant}) and $\tilde{h}(x)$ is the trace of $\tilde{h}_{ij}(x)$ under the background metric $\bar{g}_{ij}$. As a result, we indeed get a St\"{u}ckelberg mass term for the graviton with $V^{i}(x)$ as the St\"{u}ckelberg/Goldstone vector field. This completes the proof that the closed gravitational universe is in the Higgs phase, where the graviton is massive due to the spontaneously broken diffeomorphism symmetry. 

The analysis of the resulting constraint on operators follows Equ.~(\ref{eq:constraintbrane}) to Equ.~(\ref{eq:KRislando}). The nontrivial diffeomorphism invariant operators must be dressed as
\begin{equation}
    \hat{O}^{\text{Phys}}(x)=\hat{O}(x+\sqrt{16\pi G_{d}}\hat{V}(x))\,.
\end{equation}

To complete the story, we still have to prove the result corresponding to Equ.~(\ref{eq:nonlocal}), which in the current case is to compute the following commutator
\begin{equation}
    [\hat{H}_{\text{external}},\hat{O}^{\text{Phys}}(x)]\,,\label{eq:commutator}
\end{equation}
and confirm that it is not zero. For our purpose, we only need the fact that $\hat{H}_{\text{external}}$ generates the time-evolution of $\phi_{2}(x)$. The calculation of Equ.~(\ref{eq:commutator}) follows the same procedure as in Section 3.1 and Section 3.2 of \cite{Geng:2025rov},\footnote{With the minor modification incorporating the mirror charges.} which we don't repeat here. The result is simply
\begin{equation}
    [\hat{H}_{\text{external}},\hat{O}^{\text{Phys}}(x)]=i\frac{2}{d}\partial_{t}\hat{O}^{\text{Phys}}(x)\,.\label{eq:finalcommutator}
\end{equation}
We emphasize that the above nonzero result is due to the nonzero correlator $\langle T\phi_{1}(x_1)\phi_{2}(x_2)\rangle$, which indicates the entanglement between the closed universe matter field and the external non-gravitational system.

\subsection{Summary}
In this section, we established the following results:
\begin{itemize}
    \item \textbf{1)} A nontrivial Hilbert space can be associated with a closed gravitational universe if some matter fields in this closed universe are entangled with an external non-gravitational system in the $G_{N}=0$ limit.

    \item \textbf{2)} The nontrivial Hilbert space is guaranteed by the existence of nontrivial observables in this closed gravitational universe. These observables are the QFT operators dressed to the entanglement.

    \item \textbf{3)} The dressing is a gravitational effect, and these dressed operators become ordinary QFT operators in the $G_{N}=0$ limit.

    \item \textbf{4)} These dressed operators can be detected by the external non-gravitational system. This is a genuine non-local effect due to quantum mechanics and dynamical gravity.
    
    \item \textbf{5)} The mechanism of this nonlocal effect is the same as that for entanglement islands. Thus, the associated nontrivial Hilbert space should be thought of as a subspace of the Hilbert space of the external non-gravitational system. In the Karch-Randall braneworld, this is just due to the standard notion of holography.

    \item \textbf{6)} From the perspective of the closed universe, the external non-gravitational system is a ``meta-observer", and the nontrivial observables are all dressed to this ``meta-observer" by quantum entanglement.

    \item \textbf{7)} The ``meta-observer" left imprints inside the closed universe.\footnote{For example, one can compare Equ.~(\ref{eq:redef}) and Equ.~(\ref{eq:pin}).} The imprints are the canonical momenta of the Goldstone boson $\Pi_{V^{\mu}}(x)$, which by itself obeys the constraint equations. It can be described as the ``field of local observers" inside the closed gravitational universe. From this perspective, the nontrivial observables are dressed to the ``local observers" by the canonical conjugate $V^{\mu}(x)$, which can be thought of as a ``field of rods and clock".

    \item \textbf{8)} The above points are direct results of spontaneous symmetry breaking and the Higgs mechanism in the closed gravitational universe. The spontaneous symmetry breaking is induced by the quantum entanglement with the external non-gravitational system. The field $V^{\mu}(x)$ is the Goldstone boson associated with this spontaneous diffeomorphism breaking. This graviton is massive.
    
\end{itemize}

\begin{figure}
\begin{centering}
    \includegraphics[scale=0.15]{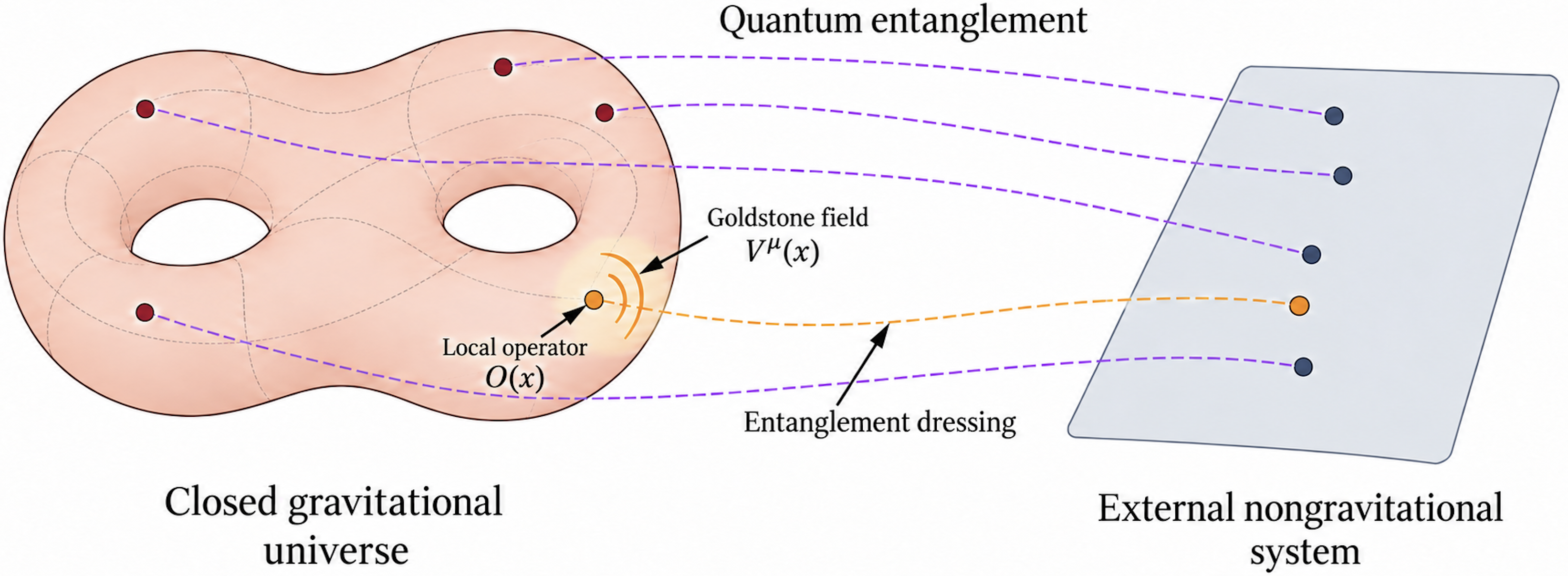}
\caption{\small  A demonstration of our result that nontrivial operators inside the gravitational closed universe are dressed to the external non-gravitational system by quantum entanglement. This figure depicts the situation at a single instant, i.e. a constant time slice. The closed universe generally has a nontrivial topology and we indicate this by drawing an example of genus-2. \label{pic:dressing}}
\end{centering}
\end{figure}

\begin{figure}
\begin{centering}
    \includegraphics[scale=0.15]{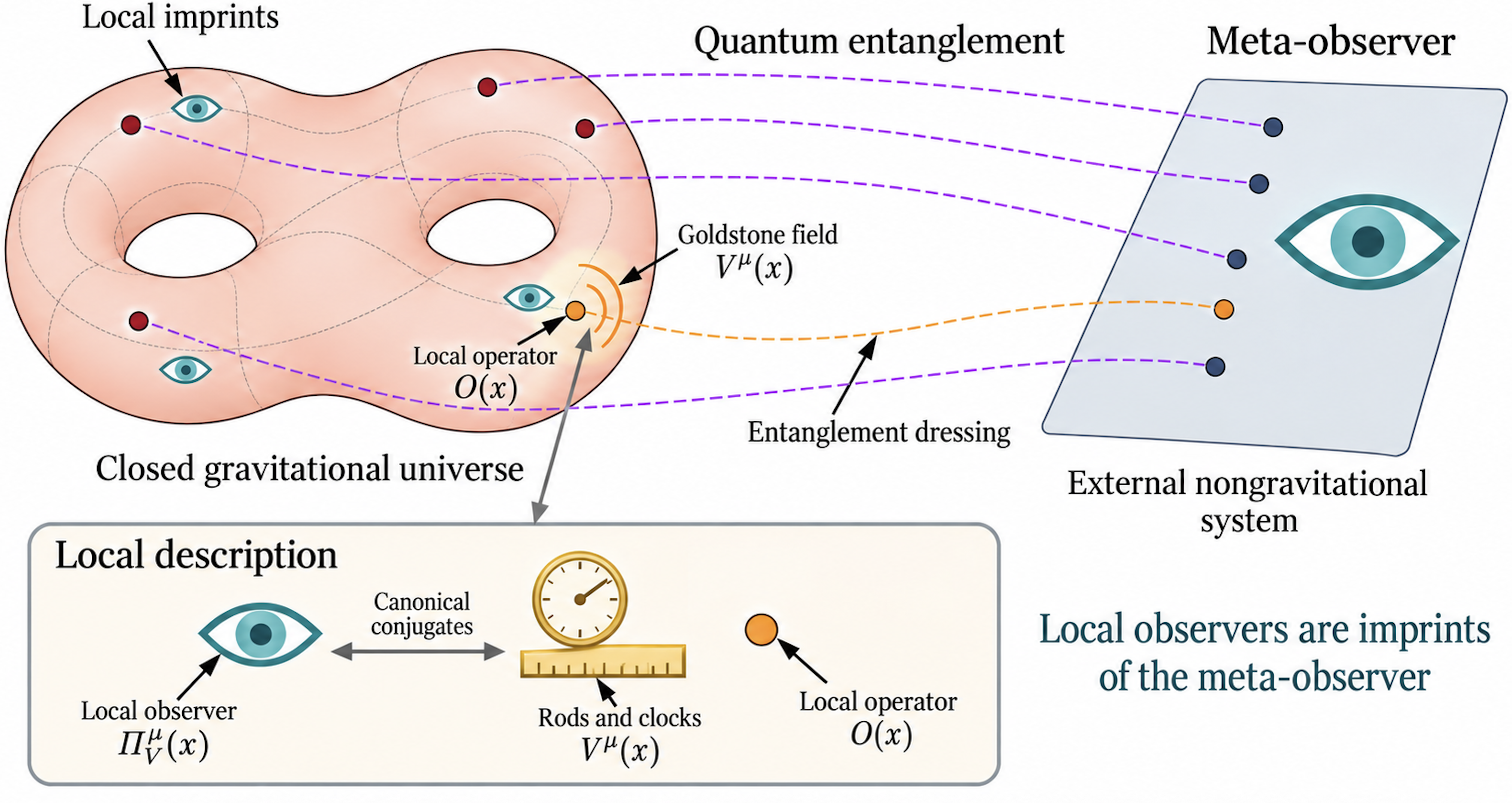}
\caption{\small  The same demonstration as Fig.~\ref{pic:dressing} in terms of ``observers". There are two types of ``observers" that we talked about. The ``meta-observer" refers to the external non-gravitational system which by holographic duality captures all the physics inside the closed gravitational universe (and more). The "local observer" inside the gravitational universe is the ``observer" that the recent literature usually talk about. Its existence ensures a nontrivial Hilbert space associated with the gravitational closed universe. From our explicit results in the Karch-Randall braneworld and that in Sec.~\ref{sec:intermediate}, the ``local observers" are in fact imprints of the ``meta-observer".\label{pic:observer}}
\end{centering}
\end{figure}

\begin{figure}
\begin{centering}
    \includegraphics[scale=0.15]{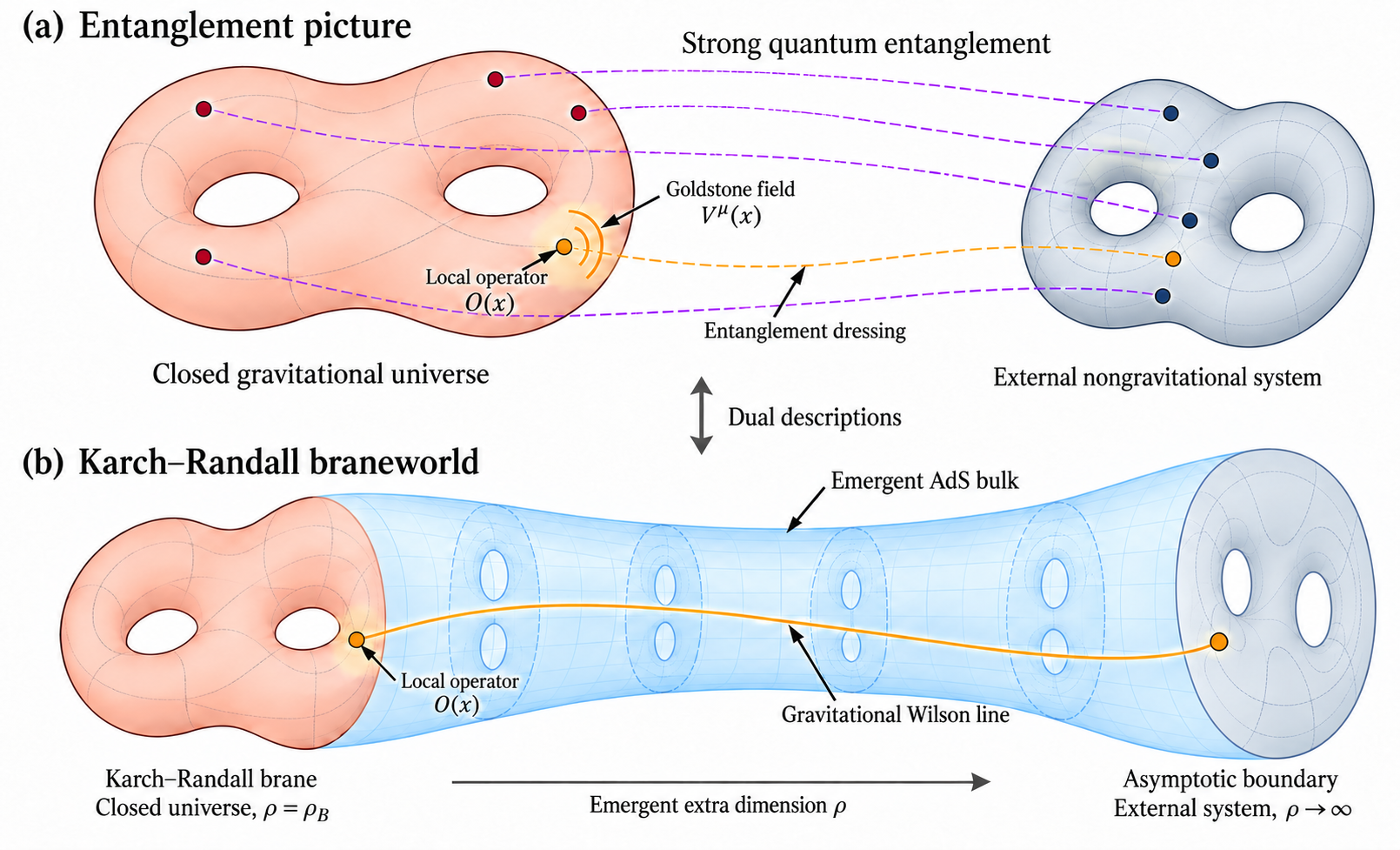}
\caption{\small The emergence of the extra-dimension due to strong entanglement. A demonstration of the Karch-Randall braneworld. \label{pic:KRemergent}}
\end{centering}
\end{figure}

The above results can be visualized in Fig.~\ref{pic:dressing} and Fig.~\ref{pic:observer}. Due to holography \cite{Geng:2025byh}, the Karch-Randall braneworld captures the situations that the entanglement between the closed gravitational universe and the external non-gravitational system is strong, i.e. a large number of entangled fields (see Fig.~\ref{pic:KRemergent}). From the intermediate description perspective, a semiclassical extra-dimension emerges as a result of this strong entanglement. This is the connection between the bulk description and the intermediate description of the Karch-Randall braneworld. Furthermore, in general cases of our construction, e.g. that in Sec.~\ref{sec:intermediate}, where the entanglement between the closed universe and the external non-gravitational system is not so strong, one can think of it as that there is a very quantum extra-dimension that connects the closed universe and the external non-gravitational system \cite{Geng:2025rov}. 

\section{Comments on the Literature and Implications}\label{sec:comments}

Having established the main technical results in this paper, let's discuss their connections to the literature and implications.

\subsection{Witten's Observer}
As it was postulated in \cite{Chandrasekaran:2022cip,Witten:2023qsv,Witten:2023xze,Jensen:2023yxy}, nontrivial observables can be constructed in a gravitational closed universe if there is an ``observer" term in the Hamiltonian
\begin{equation}
    \hat{H}_{\text{obs}}=\hat{q}\,,
\end{equation}
where $q$ is a phase space variable associated with the ``observer". The ``observer" was assumed to be localized at a particular place.\footnote{Or moving along a particular worldline.}. Thus, there exists the canonical conjugate $p$, such that
\begin{equation}
    [\hat{H}_{\text{obs}},\hat{p}]=[\hat{q},\hat{p}]=i\,.
\end{equation}
 This enables one to dress local QFT operators $\hat{O}(t)$ as
 \begin{equation}
     \hat{O}^{\text{Phys}}(t)=\hat{O}(t+\hat{p})\,,\label{eq:fake}
 \end{equation}
which obeys the integrated constraint
\begin{equation}
    [\hat{H}_{\text{QFT}}+\hat{H}_{\text{obs}},\hat{O}^{\text{phys}}(t)]=0\,.
\end{equation}

However, as it was pointed out in \cite{Geng:2024dbl,Geng:2025bcb,Geng:2025gqu,Abdalla:2025gzn}, there are a few questions associated with this proposal. First, it wasn't so clear what the microscopic origin of such an ``observer" is. There is no known closed system whose Hamiltonian is linear in its phase space variable. This suggests that such an ``observer" must be emergent from some underlying microscopic degrees of freedom or it is induced by some external system. Second, the dressed operator Equ.~(\ref{eq:fake}) is not diffeomorphism invariant, and it doesn't obey the local constraint away from the ``observer". This suggests that such a term is an oversimplified notion of a more general and realistic phenomenon that there should/could be ``observers" everywhere.

As it was realized in \cite{Geng:2024dbl,Geng:2025bcb}, the above two problems are nicely addressed in our construction in Sec.~\ref{sec:observer}. Indeed, instead of a single ``observer", we have a field of ``observers" $\Pi_{V^{\mu}}(x)$ and indeed the Hamiltonian constraint contains an extra term linear in $\Pi_{V^{0}}(x)$.\footnote{Here we use $0$ to denote the direction of the Killing time $\xi^{\mu}(x)$.} The ``observer" dressed operators $\hat{O}^{\text{Phys}}(x)=\hat{O}(x+\hat{V}(x))$ obey local constraints and are diffeomorphism invariant. Moreover, the microscopic origin of this field of ``observers" is the quantum entanglement between the matter field in the gravitational closed universe and the external non-gravitational system. More specifically, this entanglement induces the spontaneous breaking of diffeomorphisms, and the field of ``observers" emerges as the canonical momentum of the associated Goldstone vector boson. As opposed to earlier works \cite{Geng:2025byh,Geng:2023zhq,Geng:2025rov}, the construction in Sec.~\ref{sec:observer} makes it manifestly clear that the spontaneous symmetry breaking is indeed induced by quantum entanglement (more to be said in Sec.~\ref{sec:gravitonmass}). Thus, our model is an explicit realization of the proposal in \cite{Chandrasekaran:2022cip,Witten:2023qsv,Witten:2023xze,Jensen:2023yxy}, which is more general and realistic.

\subsection{Harlow-Usatyuk-Zhao Observer}\label{sec:HUZ}
Another model of ``observer" was proposed in \cite{Harlow:2025pvj}, which postulated that some quantum system $O_{b}$ inside the gravitational universe entangled with an external nongravitational system $O_{b}'$ should be identified as the ``observer". This entanglement is crucial in their quantum circuit model to ensure that the Hilbert space associated with the closed universe is not trivially one dimensional. We note that in their proposal the bulk observables are in fact encoded in the Hilbert space of $O'_{b}$. This is exactly the situation in our explicit model, where the observables in the closed gravitational universe are in fact encoded in the external non-gravitational system. Hence, our construction is an explicit physical realization of \cite{Harlow:2025pvj}. However, it deserves to be better understood if all features of their quantum circuit model are reflected in our explicit gravitational model. We leave a detailed study of this question for the future.

\subsection{Entanglement Islands and Graviton Mass}
\label{sec:gravitonmass}

As mentioned many times, the construction in this paper is highly motivated by the interior of an evaporating black hole at late-times (in Fig.~\ref{pic:interior}), which is almost a closed universe, and the fact that a nontrivial Page curve indicates nontrivial information transfer between the interior and the exterior of the black hole. Hence, the Hilbert space associated with the interior cannot be trivially one dimensional even at late-times. It was understood in \cite{Almheiri:2019psf,Penington:2019npb,Almheiri:2019yqk} that the emergence of such a nontrivial Page curve is a special case of a more general phenomenon in holography.\footnote{A relevant earlier work is \cite{Freivogel:2005qh}.} This phenomenon is the emergence of entanglement islands, which is a subregion of the gravitational universe whose physics is not encoded in the ordinary holographic dual boundary theory but in an external non-gravitational system. For the cases in \cite{Almheiri:2019psf,Penington:2019npb,Almheiri:2019yqk}, this external non-gravitational system is coupled to the gravitational universe at its boundary (as in Fig.~\ref{pic:interior}). This coupling is achieved by imposing transparent boundary condition for matter fields such that they can leak from the gravitational universe into the external system or vice versa. It was understood in \cite{Porrati:2001gx,Porrati:2003sa,Aharony:2006hz} that this coupling induces a mass for the graviton and its was argued that this mass is generated by a Higgs mechanism. This Higgs mechanism is recently understood in \cite{Geng:2023zhq,Geng:2023ynk,Geng:2025rov} which demonstrated the spontaneous diffeomorphism breaking and the emergence of the Goldstone boson. Based on these results, \cite{Geng:2020qvw} proposed/conjectured that entanglement islands should have a deep connection with the Higgs mechanism that generates the graviton mass. This proposal/conjecture was gradually understood and established in \cite{Geng:2020fxl,Geng:2021hlu,Geng:2023zhq,Geng:2025rov,Geng:2025byh}. The key observation is that the consistency of islands requires that observables inside the island should commute with the boundary ADM Hamiltonian of the gravitational universe but not commute with the Hamiltonian of the external non-gravitational system. These two conditions are obeyed by the Goldstone boson dressed operators, in the same form as Equ.~(\ref{eq:KRislando}). 

However, if one thinks more carefully one might realize the following: 1) The entanglement island is a feature of the quantum state, because it emerges from the QES formula due to the strong entanglement between the gravitational spacetime and the external non-gravitational system; 2) The graviton mass in setups like Fig.~\ref{pic:interior} is due to the coupling between the gravitational spacetime and the external non-gravitational system, so it seems to be a feature of the Hamiltonian of the system. Thus, it might be unclear if there is really a connection between the two, i.e. entanglement islands and graviton mass. This question can be formulated in a sharper way that one could decouple the gravitational spacetime and the external non-gravitational system after the island is formed. Thus, if the graviton mass is really only due to the coupling, it should become zero after this coupling. Though the island stays. Thus, there seems to be an apparent contradiction regarding the connection between islands and graviton mass. 

In fact, as it is articulated in \cite{Geng:2025byh}, the above consideration instead provides strong support for the connection between islands and graviton mass. It helps to articulate the origin of the graviton mass. As discussed, the graviton mass is a result of spontaneous diffeomorphism breaking and thus it is a feature of the quantum state as in the ordinary Higgs mechanism. More precisely, the entanglement between the gravitational spacetime and the external non-gravitational system spontaneously breaks the diffeomorphisms, and the stronger the entanglement the bigger the graviton mass. Thus, the graviton mass is really of the same origin as the islands. Nevertheless, this understanding was not supported by a clear model where the coupling is absent and the graviton is unarguably massive. Thus, our construction in Sec.~\ref{sec:observer} can be taken as such a model. The analysis in Sec.~\ref{sec:intermediate} demonstrates that the reason for the graviton mass, i.e. two copies of modes of the entangled field \cite{Geng:2023ynk,Geng:2025rov}, is really quantum entanglement, and it is simply a generalized Unruh effect.\footnote{In the Unruh effect in Minkowski space the left Rindler wedge and the right Rindler wedge are entangled. This entanglement is reflected by the fact that the left-right two point function for any QFT is not zero. This is due to the fact that the left and right Rindler wedge operators share the same set of modes, e.g. for free fields they share the same set of creation and annihilation operators. Thus, for each operator there are two sets of modes as comparing to the case when there is a reflecting boundary in between the two Rindler wedges.} A new understanding offered by this paper is that the quantum entanglement that enables islands and gives the graviton mass should be understood as a nonperturbative feature of the quantum state. This ensures that the dressed operators in the form Equ.~(\ref{eq:KRislando}) should have a nontrivial nonperturbative completion.

\subsection{Observer Counting the Wet Hair---Global Symmetries}
Having understood the underlying mechanism for the emergence of the ``observer", we can ask for its practical implications.  First, let's notice that the ``observer" field 
\begin{equation}
 \Pi^{i}_{V}(x)\equiv\bar{g}^{ij}\Pi_{V^{j}}(x)=m^2(h^{i}_{t}+\bar{\nabla}^{i}V_{t}+\bar{\nabla}_{t}V^{i})(x)\,,   
\end{equation}
is itself diffeomorphism invariant.\footnote{Remember that the combination $h_{ij}+\bar{\nabla}_{i}V_{j}+\bar{\nabla}_{j}V_{i}$ is diffeomorphism invariant.} Therefore, the ``observer" field is physical. Moreover, the ``observers" described by the ``observer" field can measure the observables $\hat{O}^{\text{Phys}}(x)$. This is because an equal-time measurement will give
\begin{equation}
\begin{split}
    [\Pi^{i}_{V}(x),\hat{O}^{\text{Phys}}(y)]&= [\Pi^{i}_{V}(x),\hat{O}(xy+\sqrt{16\pi G_{N}}\hat{V}(y))]\\&=-i\sqrt{16\pi G_{N}}\bar{g}^{ij}\partial_{j}\hat{O}^{\text{Phys}}(x)\frac{1}{\sqrt{-\bar{g}}}\delta^{d-1}(\vec{x}-\vec{y})\,.\label{eq:observerobs}
    \end{split}
\end{equation}
In other words, the ``observers" could measure the ``energy" and ``momentum" sourced by the dressed observables. This is a rather surprising implication, as even though we know that in a gravitational universe there is no local notion of energy and momentum, the ``observers" have a notion of energy and momentum, i.e. locations and time. From Equ.~(\ref{eq:nonlocal}) and Equ.~(\ref{eq:finalcommutator}), one can interpret this notion of energy and momentum as an inheritance or imprint from the external non-gravitational system, which does have a well-defined notion of energy and momentum. From the perspective of the Karch-Randall braneworld, the ``observers" could measure energy and momentum by ``counting" the local number of gravitational flux lines (Wilson lines) that disappear into the extra-dimension.

This observation easily generalizes to other global symmetries, i.e., conserved charges.\footnote{Actually, in the above discussion, the conserved CFT Hamiltonian should be generating the conformal Killing time, i.e., $\eta$, translation.} The essential generalization is to consider almost the same story as in Sec.~\ref{sec:observer}, but instead of the graviton one considers gauge fields in the closed universe, and the matter fields that are entangled with an external non-gravitational system are charged under the gauge symmetry. Moreover, the external system has a global symmetry, which is the same symmetry group as the gauge symmetry in the gravitational system, and the entangled matter fields are charged under this global symmetry. Then everything in Sec.~\ref{sec:observer} goes through. We refer the readers to \cite{Geng:2025gns} for details. The conclusions are that the gauge symmetry in the gravitational system is spontaneously broken by quantum entanglement, the gauge boson becomes massive, and there is an emergent Goldstone boson inside the gravitational closed universe, which however has a nonzero commutator with the global symmetry charge of the external non-gravitational system. Furthermore, similar to the Hamiltonian constraint, we also have to impose the gauge constraint on observables inside the closed gravitational universe. The nontrivial charged operators inside the closed gravitational universe should be dressed by the Goldstone boson in the same manner as Equ.~(\ref{eq:KRislando}), which as a result is charged under the global symmetry of the external non-gravitational system. Thus, there could exist global symmetry charged operators inside the gravitational closed universe, and also entanglement islands \cite{Geng:2025gns}, i.e. the black hole interior. This is called ``Wet Hair" in \cite{Geng:2025gns}. The ``observers" are now equipped with another gauge invariant operator--- the canonical conjugate of the Goldstone boson, and they therefore have a notion of global symmetry in the same manner as Equ.~(\ref{eq:observerobs}). Similar to before, the ``observers" are essentially measuring the global symmetry charge by ``counting" the number of gauge flux lines that disappears into the extra-dimension, i.e. ``counting" the number of wet hair!

\subsection{Quantum Entanglement and the Emergence of Spacetime and Gravity}

As it was pointed out in \cite{Geng:2025rov}, the Karch-Randall braneworld should be taken as an explicit example that gravitational spacetime could emerge from strong quantum entanglement (see Fig.~\ref{pic:KRemergent}). This observation is recently used in \cite{Geng:2025efs} to provide an explicit realization of the emergence of a gravitational AdS$_{3}$ from a CFT$_{2}$ with a large central charge and a sparse low-lying spectrum. The basic idea is to cut the CFT state preparing path integral using many small-hole Cardy boundaries and then derive the emergence of a gravitational AdS$_{3}$ with many small pieces of Karch-Randall branes. More interestingly, the spacetime between nearby small pieces of the Karch-Randall branes is in fact a black hole, so the full emergent AdS$_{3}$ spacetime can be thought of as a condensation of these tiny black holes. These tiny black holes are strongly entangled with each other, and so the emergence of the AdS$_{3}$ is really a result of quantum entanglement. This is an explicit realization of the ER=EPR proposal \cite{VanRaamsdonk:2010pw,Maldacena:2013xja}

We believe that it is an interesting question to extend the techniques in \cite{Geng:2025efs,Bao:2025plr,Bao:2024ixc,Chen:2024unp,Hung:2024gma,Chua:2023ios} to the current context. For example, how to derive the existence of a closed universe directly from the external non-gravitational system. We leave this question for the future.

\section{Conclusions and Discussions}\label{sec:conclusion}
In this paper, we constructed an explicit model of a closed gravitational universe with ``observers" and as a result a nontrivial Hilbert space. This construction follows the general idea in \cite{Geng:2024dbl,Geng:2025gqu}, where the ``observers" emerge from spontaneous diffeomorphism breaking. The construction in this paper provided an explicit realization of the proposal in \cite{Harlow:2025pvj} and unified it with the earlier works \cite{Chandrasekaran:2022cip,Witten:2023qsv,Witten:2023xze}, with both as special cases of \cite{Geng:2024dbl,Geng:2025gqu}. The construction is motivated by the physics of evaporating black holes and entanglement islands. The essential observation is that the interiors of evaporating black holes at later-times are (almost) closed universes and at the same time they are entanglement islands of the early Hawking-radiation. Thus, the physics inside the closed gravitational universe should be the same as that of entanglement islands. Our construction explicitly demonstrated this observation. The earlier observation that entanglement islands exist only in massive gravity and the graviton mass is essential in ensuring the consistency of the islands \cite{Geng:2020qvw,Geng:2020fxl,Geng:2021hlu,Geng:2025rov,Geng:2025byh,Geng:2025byh} plays a crucial role in understanding the emergence of the ``observers" and the nontrivial Hilbert space associated with the closed gravitational universe. From this perspective, the ``observer" paradigm is a rephrasing of the \textit{Massive Islands} paradigm. Our construction provides a clear example to demonstrate that the graviton mass is a feature of the quantum state so as entanglement islands. An interesting implication of our construction is that the ``observers" have notions of global symmetries. Therefore, these ``observers" sound quite like ourselves, but we leave a more careful study of this question for the future.

\section*{Acknowledgments}
We thank Ahmed Almheiri, Stefano Antonini, Gabriel Arenas-Henriquez, Liam A. Fitzpatrick, \AA smund Folksted, Steve Giddings, Daniel Harlow, Tom Hartman, Ling-Yan Hung, Daniel Jafferis, Yikun Jiang, Andreas Karch, Emanuel Katz, Henry W. Lin, Jacob McNamara, Juan Maldacena, Henry Maxfield, Rong-Xin Miao, Rashmish Mishra, Masamichi Miyaji, Manqian Ou, Geoff Penington, Carlos Perez-Pardavila, Suvrat Raju, Lisa Randall, Matt Reece, Martin Sasieta, Pushkal Shrivastava, Jia Tian, Huy Tran, Jiuci Xu, Zhenbin Yang and Yoav Zigdon for helpful discussions. We thank Jes\'{u}s Huertas, Ling-Yan Hung, Yikun Jiang, Andreas Karch, Daniel Jaffeirs, Moritz Merz,Lisa Randall, Neeraj Tata, Pushkal Shrivastava and Dawson Thomas for relevant collaborations. We thank the authors of \cite{Antonini:2025sur} (Stefano Antonini, Chang-Han Chen, Henry Maxfield and Geoff Penington) for sharing the prime version of their draft and Henry Maxfield and Geoff Penington for the correspondence about the prime version of \cite{Antonini:2025sur}. This paper comes from illuminating questions we have heard from many people over the last few years. We also thank ChatGPT-6 Astra for the help in drawing Fig.~\ref{pic:dressing}, Fig.~\ref{pic:observer}, and Fig.~\ref{pic:KRemergent}. I also use Claude to check the calculations in this paper. HG is supported by the Gravity, Spacetime, and Particle Physics (GRASP) Initiative from Harvard University and a grant from the Physics Department at Harvard University. 
\bibliographystyle{JHEP}

\bibliography{main}

@article{Almheiri:2019psy,
    author = "Almheiri, Ahmed and Mahajan, Raghu and Santos, Jorge E.",
    title = "{Entanglement islands in higher dimensions}",
    eprint = "1911.09666",
    archivePrefix = "arXiv",
    primaryClass = "hep-th",
    doi = "10.21468/SciPostPhys.9.1.001",
    journal = "SciPost Phys.",
    volume = "9",
    number = "1",
    pages = "001",
    year = "2020"
}

@article{Geng:2021hlu,
    author = "Geng, Hao and Karch, Andreas and Perez-Pardavila, Carlos and Raju, Suvrat and Randall, Lisa and Riojas, Marcos and Shashi, Sanjit",
    title = "{Inconsistency of Islands in Theories with Long-Range Gravity}",
    eprint = "2107.03390",
    archivePrefix = "arXiv",
    primaryClass = "hep-th",
    month = "7",
    year = "2021"
}

@article{Porrati:2001gx,
    author = "Porrati, Massimo",
    title = "{Mass and gauge invariance 4. Holography for the Karch-Randall model}",
    eprint = "hep-th/0109017",
    archivePrefix = "arXiv",
    reportNumber = "NYU-TH-01-08-02",
    doi = "10.1103/PhysRevD.65.044015",
    journal = "Phys. Rev. D",
    volume = "65",
    pages = "044015",
    year = "2002"
}

@article{Karch:2000ct,
    author = "Karch, Andreas and Randall, Lisa",
    editor = "Duff, Michael J. and Liu, J.T. and Lu, J.",
    title = "{Locally localized gravity}",
    eprint = "hep-th/0011156",
    archivePrefix = "arXiv",
    reportNumber = "MIT-CTP-3099",
    doi = "10.1088/1126-6708/2001/05/008",
    journal = "JHEP",
    volume = "05",
    pages = "008",
    year = "2001"
}

@article{deRham:2014zqa,
    author = "de Rham, Claudia",
    title = "{Massive Gravity}",
    eprint = "1401.4173",
    archivePrefix = "arXiv",
    primaryClass = "hep-th",
    doi = "10.12942/lrr-2014-7",
    journal = "Living Rev. Rel.",
    volume = "17",
    pages = "7",
    year = "2014"
}

@inproceedings{deRham:2026pxl,
    author = "de Rham, Claudia",
    title = "{Massive Gravity @ 15}",
    eprint = "2607.15507",
    archivePrefix = "arXiv",
    primaryClass = "hep-th",
    reportNumber = "Imperial--TP--2026--cdr--3",
    month = "7",
    year = "2026"
}

@book{dirac2001quantum,
  title={Lectures on Quantum mechanics},
  author={Dirac, Paul Adrien Maurice},
  year={2001},
  publisher={Dover},
  note = {Original published by Belfer Graduate School of Science in 1964},
}

@article{Almheiri:2019psf,
    author = "Almheiri, Ahmed and Engelhardt, Netta and Marolf, Donald and Maxfield, Henry",
    title = "{The entropy of bulk quantum fields and the entanglement wedge of an evaporating black hole}",
    eprint = "1905.08762",
    archivePrefix = "arXiv",
    primaryClass = "hep-th",
    doi = "10.1007/JHEP12(2019)063",
    journal = "JHEP",
    volume = "12",
    pages = "063",
    year = "2019"
}

@article{Laddha:2020kvp,
    author = "Laddha, Alok and Prabhu, Siddharth G. and Raju, Suvrat and Shrivastava, Pushkal",
    title = "{The Holographic Nature of Null Infinity}",
    eprint = "2002.02448",
    archivePrefix = "arXiv",
    primaryClass = "hep-th",
    doi = "10.21468/SciPostPhys.10.2.041",
    journal = "SciPost Phys.",
    volume = "10",
    pages = "041",
    year = "2021"
}

@article{Donnelly:2015hta,
      author         = "Donnelly, William and Giddings, Steven B.",
      title          = "{Diffeomorphism-invariant observables and their nonlocal
                        algebra}",
      journal        = "Phys. Rev.",
      volume         = "D93",
      year           = "2016",
      number         = "2",
      pages          = "024030",
      doi            = "10.1103/PhysRevD.94.029903, 10.1103/PhysRevD.93.024030",
      note           = "[Erratum: Phys. Rev.D94,no.2,029903(2016)]",
      eprint         = "1507.07921",
      archivePrefix  = "arXiv",
      primaryClass   = "hep-th",
      reportNumber   = "NSF-KITP-15-133",
      SLACcitation   = "%%CITATION = ARXIV:1507.07921;%%"
}

@article{Almheiri:2019qdq,
    author = "Almheiri, Ahmed and Hartman, Thomas and Maldacena, Juan and Shaghoulian, Edgar and Tajdini, Amirhossein",
    title = "{Replica Wormholes and the Entropy of Hawking Radiation}",
    eprint = "1911.12333",
    archivePrefix = "arXiv",
    primaryClass = "hep-th",
    doi = "10.1007/JHEP05(2020)013",
    journal = "JHEP",
    volume = "05",
    pages = "013",
    year = "2020"
}

@book{peskin1995iqf,
  title={{An Introduction to Quantum Field Theory}},
  author={Peskin, M.E. and Schroeder, D.V.},
  year={1995},
  publisher={Perseus Books}
}

@article{Chen:2020hmv,
    author = "Chen, Hong Zhe and Myers, Robert C. and Neuenfeld, Dominik and Reyes, Ignacio A. and Sandor, Joshua",
    title = "{Quantum Extremal Islands Made Easy, Part II: Black Holes on the Brane}",
    eprint = "2010.00018",
    archivePrefix = "arXiv",
    primaryClass = "hep-th",
    month = "9",
    year = "2020"
}

@article{Donnelly:2016rvo,
      author         = "Donnelly, William and Giddings, Steven B.",
      title          = "{Observables, gravitational dressing, and obstructions to
                        locality and subsystems}",
      journal        = "Phys. Rev.",
      volume         = "D94",
      year           = "2016",
      number         = "10",
      pages          = "104038",
      doi            = "10.1103/PhysRevD.94.104038",
      eprint         = "1607.01025",
      archivePrefix  = "arXiv",
      primaryClass   = "hep-th",
      SLACcitation   = "%%CITATION = ARXIV:1607.01025;%%"
}

@article{Geng:2020fxl,
    author = "Geng, Hao and Karch, Andreas and Perez-Pardavila, Carlos and Raju, Suvrat and Randall, Lisa and Riojas, Marcos and Shashi, Sanjit",
    title = "{Information Transfer with a Gravitating Bath}",
    eprint = "2012.04671",
    archivePrefix = "arXiv",
    primaryClass = "hep-th",
    month = "12",
    year = "2020"
}

@article{Raju:2018zpn,
    author = "Raju, Suvrat",
    title = "{A Toy Model of the Information Paradox in Empty Space}",
    eprint = "1809.10154",
    archivePrefix = "arXiv",
    primaryClass = "hep-th",
    doi = "10.21468/SciPostPhys.6.6.073",
    journal = "SciPost Phys.",
    volume = "6",
    number = "6",
    pages = "073",
    year = "2019"
}

@article{Almheiri:2019hni,
    author = "Almheiri, Ahmed and Mahajan, Raghu and Maldacena, Juan and Zhao, Ying",
    title = "{The Page curve of Hawking radiation from semiclassical geometry}",
    eprint = "1908.10996",
    archivePrefix = "arXiv",
    primaryClass = "hep-th",
    doi = "10.1007/JHEP03(2020)149",
    journal = "JHEP",
    volume = "03",
    pages = "149",
    year = "2020"
}

@article{Lashkari:2013koa,
      author         = "Lashkari, Nima and McDermott, Michael B. and Van
                        Raamsdonk, Mark",
      title          = "{Gravitational dynamics from entanglement
                        'thermodynamics'}",
      journal        = "JHEP",
      volume         = "1404",
      pages          = "195",
      doi            = "10.1007/JHEP04(2014)195",
      year           = "2014",
      eprint         = "1308.3716",
      archivePrefix  = "arXiv",
      primaryClass   = "hep-th",
      SLACcitation   = "%%CITATION = ARXIV:1308.3716;%%",
}

@article{Takayanagi:2011zk,
    author = "Takayanagi, Tadashi",
    title = "{Holographic Dual of BCFT}",
    eprint = "1105.5165",
    archivePrefix = "arXiv",
    primaryClass = "hep-th",
    reportNumber = "IPMU11-0091",
    doi = "10.1103/PhysRevLett.107.101602",
    journal = "Phys. Rev. Lett.",
    volume = "107",
    pages = "101602",
    year = "2011"
}

@article{Giddings:2019hjc,
    author = "Giddings, Steven B.",
    title = "{Gravitational dressing, soft charges, and perturbative gravitational splitting}",
    eprint = "1903.06160",
    archivePrefix = "arXiv",
    primaryClass = "hep-th",
    reportNumber = "CERN-TH-2019-074",
    doi = "10.1103/PhysRevD.100.126001",
    journal = "Phys. Rev. D",
    volume = "100",
    number = "12",
    pages = "126001",
    year = "2019"
}

@article{Aharony:2006hz,
    author = "Aharony, Ofer and Clark, Adam B. and Karch, Andreas",
    title = "{The CFT/AdS correspondence, massive gravitons and a connectivity index conjecture}",
    eprint = "hep-th/0608089",
    archivePrefix = "arXiv",
    reportNumber = "WIS-09-06-AUG-DPP",
    doi = "10.1103/PhysRevD.74.086006",
    journal = "Phys. Rev. D",
    volume = "74",
    pages = "086006",
    year = "2006"
}

@article{Penington:2019kki,
    author = "Penington, Geoff and Shenker, Stephen H. and Stanford, Douglas and Yang, Zhenbin",
    title = "{Replica wormholes and the black hole interior}",
    eprint = "1911.11977",
    archivePrefix = "arXiv",
    primaryClass = "hep-th",
    month = "11",
    year = "2019"
}

@article{Kourkoulou:2017zaj,
    author = "Kourkoulou, Ioanna and Maldacena, Juan",
    title = "{Pure states in the SYK model and nearly-$AdS_2$ gravity}",
    eprint = "1707.02325",
    archivePrefix = "arXiv",
    primaryClass = "hep-th",
    month = "7",
    year = "2017"
}

@article{Porrati:2003sa,
    author = "Porrati, M.",
    title = "{Higgs phenomenon for the graviton in ADS space}",
    eprint = "hep-th/0306253",
    archivePrefix = "arXiv",
    doi = "10.1142/S0217732303011745",
    journal = "Mod. Phys. Lett. A",
    volume = "18",
    pages = "1793--1802",
    year = "2003"
}

@article{Fujita:2011fp,
    author = "Fujita, Mitsutoshi and Takayanagi, Tadashi and Tonni, Erik",
    title = "{Aspects of AdS/BCFT}",
    eprint = "1108.5152",
    archivePrefix = "arXiv",
    primaryClass = "hep-th",
    reportNumber = "IPMU-11-0136, MIT-CTP-4289",
    doi = "10.1007/JHEP11(2011)043",
    journal = "JHEP",
    volume = "11",
    pages = "043",
    year = "2011"
}

@article{Maldacena:1997re,
    author = "Maldacena, Juan Martin",
    title = "{The Large N limit of superconformal field theories and supergravity}",
    eprint = "hep-th/9711200",
    archivePrefix = "arXiv",
    reportNumber = "HUTP-97-A097, HUTP-98-A097",
    doi = "10.1023/A:1026654312961",
    journal = "Int. J. Theor. Phys.",
    volume = "38",
    pages = "1113--1133",
    year = "1999"
}

@article{Almheiri:2012rt,
    author = "Almheiri, Ahmed and Marolf, Donald and Polchinski, Joseph and Sully, James",
    title = "{Black Holes: Complementarity or Firewalls?}",
    eprint = "1207.3123",
    archivePrefix = "arXiv",
    primaryClass = "hep-th",
    doi = "10.1007/JHEP02(2013)062",
    journal = "JHEP",
    volume = "02",
    pages = "062",
    year = "2013"
}

@article{Witten:1998qj,
    author = "Witten, Edward",
    title = "{Anti-de Sitter space and holography}",
    eprint = "hep-th/9802150",
    archivePrefix = "arXiv",
    reportNumber = "IASSNS-HEP-98-15",
    doi = "10.4310/ATMP.1998.v2.n2.a2",
    journal = "Adv. Theor. Math. Phys.",
    volume = "2",
    pages = "253--291",
    year = "1998"
}

@article{Mathur:2008nj,
      author         = "Mathur, Samir D.",
      title          = "{Fuzzballs and the information paradox: A Summary and
                        conjectures}",
      year           = "2008",
      eprint         = "0810.4525",
      archivePrefix  = "arXiv",
      primaryClass   = "hep-th",
      SLACcitation   = "%%CITATION = ARXIV:0810.4525;%%"
}

@article{Almheiri:2019yqk,
    author = "Almheiri, Ahmed and Mahajan, Raghu and Maldacena, Juan",
    title = "{Islands outside the horizon}",
    eprint = "1910.11077",
    archivePrefix = "arXiv",
    primaryClass = "hep-th",
    month = "10",
    year = "2019"
}

@article{VanRaamsdonk:2010pw,
    author = "Van Raamsdonk, Mark",
    title = "{Building up spacetime with quantum entanglement}",
    eprint = "1005.3035",
    archivePrefix = "arXiv",
    primaryClass = "hep-th",
    doi = "10.1142/S0218271810018529",
    journal = "Gen. Rel. Grav.",
    volume = "42",
    pages = "2323--2329",
    year = "2010"
}

@article{Hawking:1976ra,
    author = "Hawking, S.W.",
    title = "{Breakdown of Predictability in Gravitational Collapse}",
    doi = "10.1103/PhysRevD.14.2460",
    journal = "Phys. Rev. D",
    volume = "14",
    pages = "2460--2473",
    year = "1976"
}

@article{Cooper:2018cmb,
    author = "Cooper, Sean and Rozali, Moshe and Swingle, Brian and Van Raamsdonk, Mark and Waddell, Christopher and Wakeham, David",
    title = "{Black hole microstate cosmology}",
    eprint = "1810.10601",
    archivePrefix = "arXiv",
    primaryClass = "hep-th",
    doi = "10.1007/JHEP07(2019)065",
    journal = "JHEP",
    volume = "07",
    pages = "065",
    year = "2019"
}

@article{Penington:2019npb,
    author = "Penington, Geoffrey",
    title = "{Entanglement Wedge Reconstruction and the Information Paradox}",
    eprint = "1905.08255",
    archivePrefix = "arXiv",
    primaryClass = "hep-th",
    doi = "10.1007/JHEP09(2020)002",
    journal = "JHEP",
    volume = "09",
    pages = "002",
    year = "2020"
}

@article{Freivogel:2005qh,
    author = "Freivogel, Ben and Hubeny, Veronika E. and Maloney, Alexander and Myers, Robert C. and Rangamani, Mukund and Shenker, Stephen",
    title = "{Inflation in AdS/CFT}",
    eprint = "hep-th/0510046",
    archivePrefix = "arXiv",
    reportNumber = "SLAC-PUB-11505, SU-ITP-05-27, UCB-PTH-05-30, LBNL-58913, DCPT-05-45",
    doi = "10.1088/1126-6708/2006/03/007",
    journal = "JHEP",
    volume = "03",
    pages = "007",
    year = "2006"
}

@article{Hinterbichler:2011tt,
    author = "Hinterbichler, Kurt",
    title = "{Theoretical Aspects of Massive Gravity}",
    eprint = "1105.3735",
    archivePrefix = "arXiv",
    primaryClass = "hep-th",
    doi = "10.1103/RevModPhys.84.671",
    journal = "Rev. Mod. Phys.",
    volume = "84",
    pages = "671--710",
    year = "2012"
}

@article{Maldacena:2013xja,
    author = "Maldacena, Juan and Susskind, Leonard",
    title = "{Cool horizons for entangled black holes}",
    eprint = "1306.0533",
    archivePrefix = "arXiv",
    primaryClass = "hep-th",
    doi = "10.1002/prop.201300020",
    journal = "Fortsch. Phys.",
    volume = "61",
    pages = "781--811",
    year = "2013"
}

@article{Papadodimas:2013jku,
    author = "Papadodimas, Kyriakos and Raju, Suvrat",
    title = "{State-Dependent Bulk-Boundary Maps and Black Hole Complementarity}",
    eprint = "1310.6335",
    archivePrefix = "arXiv",
    primaryClass = "hep-th",
    reportNumber = "ICTS-2013-21",
    doi = "10.1103/PhysRevD.89.086010",
    journal = "Phys. Rev. D",
    volume = "89",
    number = "8",
    pages = "086010",
    year = "2014"
}

@article{Gubser:1998bc,
    author = "Gubser, S.S. and Klebanov, Igor R. and Polyakov, Alexander M.",
    title = "{Gauge theory correlators from noncritical string theory}",
    eprint = "hep-th/9802109",
    archivePrefix = "arXiv",
    reportNumber = "PUPT-1767",
    doi = "10.1016/S0370-2693(98)00377-3",
    journal = "Phys. Lett. B",
    volume = "428",
    pages = "105--114",
    year = "1998"
}

@article{Mathur:2009hf,
    author = "Mathur, Samir D.",
    editor = "Uranga, A.M.",
    title = "{The Information paradox: A Pedagogical introduction}",
    eprint = "0909.1038",
    archivePrefix = "arXiv",
    primaryClass = "hep-th",
    doi = "10.1088/0264-9381/26/22/224001",
    journal = "Class. Quant. Grav.",
    volume = "26",
    pages = "224001",
    year = "2009"
}

@article{Arnowitt:1962hi,
    author = "Arnowitt, Richard L. and Deser, Stanley and Misner, Charles W.",
    title = "{The Dynamics of general relativity}",
    eprint = "gr-qc/0405109",
    archivePrefix = "arXiv",
    doi = "10.1007/s10714-008-0661-1",
    journal = "Gen. Rel. Grav.",
    volume = "40",
    pages = "1997--2027",
    year = "2008"
}

@article{Raju:2020smc,
    author = "Raju, Suvrat",
    title = "{Lessons from the Information Paradox}",
    eprint = "2012.05770",
    archivePrefix = "arXiv",
    primaryClass = "hep-th",
    month = "12",
    year = "2020"
}

@article{Papadodimas:2012aq,
    author = "Papadodimas, Kyriakos and Raju, Suvrat",
    title = "{An Infalling Observer in AdS/CFT}",
    eprint = "1211.6767",
    archivePrefix = "arXiv",
    primaryClass = "hep-th",
    reportNumber = "HRI-ST-1107, ICTS-2012-10",
    doi = "10.1007/JHEP10(2013)212",
    journal = "JHEP",
    volume = "10",
    pages = "212",
    year = "2013"
}

@article{Geng:2024xpj,
    author = "Geng, Hao",
    title = "{Replica wormholes and entanglement islands in the Karch-Randall braneworld}",
    eprint = "2405.14872",
    archivePrefix = "arXiv",
    primaryClass = "hep-th",
    doi = "10.1007/JHEP01(2025)063",
    journal = "JHEP",
    volume = "01",
    pages = "063",
    year = "2025"
}

@article{Almheiri:2013hfa,
    author = "Almheiri, Ahmed and Marolf, Donald and Polchinski, Joseph and Stanford, Douglas and Sully, James",
    title = "{An Apologia for Firewalls}",
    eprint = "1304.6483",
    archivePrefix = "arXiv",
    primaryClass = "hep-th",
    doi = "10.1007/JHEP09(2013)018",
    journal = "JHEP",
    volume = "09",
    pages = "018",
    year = "2013"
}

@article{Karch:2000gx,
    author = "Karch, Andreas and Randall, Lisa",
    title = "{Open and closed string interpretation of SUSY CFT's on branes with boundaries}",
    eprint = "hep-th/0105132",
    archivePrefix = "arXiv",
    reportNumber = "MIT-CTP-3146",
    doi = "10.1088/1126-6708/2001/06/063",
    journal = "JHEP",
    volume = "06",
    pages = "063",
    year = "2001"
}

@article{Geng:2020qvw,
    author = "Geng, Hao and Karch, Andreas",
    title = "{Massive islands}",
    eprint = "2006.02438",
    archivePrefix = "arXiv",
    primaryClass = "hep-th",
    doi = "10.1007/JHEP09(2020)121",
    journal = "JHEP",
    volume = "09",
    pages = "121",
    year = "2020"
}

@article{Antonini:2021xar,
    author = "Antonini, Stefano and Swingle, Brian",
    title = "{Holographic boundary states and dimensionally-reduced braneworld spacetimes}",
    eprint = "2105.02912",
    archivePrefix = "arXiv",
    primaryClass = "hep-th",
    month = "5",
    year = "2021"
}

@article{Hawking:1974sw,
    author = "Hawking, S.W.",
    editor = "Gibbons, G.W. and Hawking, S.W.",
    title = "{Particle Creation by Black Holes}",
    doi = "10.1007/BF02345020",
    journal = "Commun. Math. Phys.",
    volume = "43",
    pages = "199--220",
    year = "1975",
    note = "[Erratum: Commun.Math.Phys. 46, 206 (1976)]"
}

@article{Chen:2020uac,
    author = "Chen, Hong Zhe and Myers, Robert C. and Neuenfeld, Dominik and Reyes, Ignacio A. and Sandor, Joshua",
    title = "{Quantum Extremal Islands Made Easy, Part I: Entanglement on the Brane}",
    eprint = "2006.04851",
    archivePrefix = "arXiv",
    primaryClass = "hep-th",
    doi = "10.1007/JHEP10(2020)166",
    journal = "JHEP",
    volume = "10",
    pages = "166",
    year = "2020"
}

@article{Raju:2021lwh,
    author = "Raju, Suvrat",
    title = "{Failure of the split property in gravity and the information paradox}",
    eprint = "2110.05470",
    archivePrefix = "arXiv",
    primaryClass = "hep-th",
    doi = "10.1088/1361-6382/ac482b",
    journal = "Class. Quant. Grav.",
    volume = "39",
    number = "6",
    pages = "064002",
    year = "2022"
}

@article{Chakraborty:2023los,
    author = "Chakraborty, Tuneer and Chakravarty, Joydeep and Godet, Victor and Paul, Priyadarshi and Raju, Suvrat",
    title = "{Holography of information in de Sitter space}",
    eprint = "2303.16316",
    archivePrefix = "arXiv",
    primaryClass = "hep-th",
    month = "3",
    year = "2023"
}

@article{Geng:2023zhq,
    author = "Geng, Hao",
    title = "{Graviton Mass and Entanglement Islands in Low Spacetime Dimensions}",
    eprint = "2312.13336",
    archivePrefix = "arXiv",
    primaryClass = "hep-th",
    month = "12",
    year = "2023"
}

@article{Geng:2023ynk,
    author = "Geng, Hao",
    title = "{Open AdS/CFT via a Double Trace Deformation}",
    eprint = "2311.13633",
    archivePrefix = "arXiv",
    primaryClass = "hep-th",
    month = "11",
    year = "2023"
}

@article{Witten:2023xze,
    author = "Witten, Edward",
    title = "{A Background Independent Algebra in Quantum Gravity}",
    eprint = "2308.03663",
    archivePrefix = "arXiv",
    primaryClass = "hep-th",
    month = "8",
    year = "2023"
}

@article{Witten:2023qsv,
    author = "Witten, Edward",
    title = "{Algebras, Regions, and Observers}",
    eprint = "2303.02837",
    archivePrefix = "arXiv",
    primaryClass = "hep-th",
    month = "3",
    year = "2023"
}

@article{Chandrasekaran:2022cip,
    author = "Chandrasekaran, Venkatesa and Longo, Roberto and Penington, Geoff and Witten, Edward",
    title = "{An algebra of observables for de Sitter space}",
    eprint = "2206.10780",
    archivePrefix = "arXiv",
    primaryClass = "hep-th",
    doi = "10.1007/JHEP02(2023)082",
    journal = "JHEP",
    volume = "02",
    pages = "082",
    year = "2023"
}

@article{Jensen:2023yxy,
    author = "Jensen, Kristan and Sorce, Jonathan and Speranza, Antony J.",
    title = "{Generalized entropy for general subregions in quantum gravity}",
    eprint = "2306.01837",
    archivePrefix = "arXiv",
    primaryClass = "hep-th",
    doi = "10.1007/JHEP12(2023)020",
    journal = "JHEP",
    volume = "12",
    pages = "020",
    year = "2023"
}

@article{Nomura:2026igt,
    author = "Nomura, Yasunori and Ugajin, Tomonori",
    title = "{Physical Predictions in Closed Quantum Gravity}",
    eprint = "2602.13387",
    archivePrefix = "arXiv",
    primaryClass = "hep-th",
    reportNumber = "RIKEN-iTHEMS-Report-26",
    month = "2",
    year = "2026"
}

@article{Balasubramanian:2023xyd,
    author = "Balasubramanian, Vijay and Nomura, Yasunori and Ugajin, Tomonori",
    title = "{De Sitter space is sometimes not empty}",
    eprint = "2308.09748",
    archivePrefix = "arXiv",
    primaryClass = "hep-th",
    reportNumber = "RIKEN-iTHEMS-Report-23",
    doi = "10.1007/JHEP02(2024)135",
    journal = "JHEP",
    volume = "02",
    pages = "135",
    year = "2024"
}

@article{Maldacena:2024spf,
    author = "Maldacena, Juan",
    title = "{Real observers solving imaginary problems}",
    eprint = "2412.14014",
    archivePrefix = "arXiv",
    primaryClass = "hep-th",
    month = "12",
    year = "2024"
}

@article{Geng:2023qwm,
    author = "Geng, Hao",
    title = "{Revisiting Recent Progress in the Karch-Randall Braneworld}",
    eprint = "2306.15671",
    archivePrefix = "arXiv",
    primaryClass = "hep-th",
    month = "6",
    year = "2023"
}

@article{Giddings:2018umg,
    author = "Giddings, Steven B. and Kinsella, Alex",
    title = "{Gauge-invariant observables, gravitational dressings, and holography in AdS}",
    eprint = "1802.01602",
    archivePrefix = "arXiv",
    primaryClass = "hep-th",
    doi = "10.1007/JHEP11(2018)074",
    journal = "JHEP",
    volume = "11",
    pages = "074",
    year = "2018"
}

@article{Geng:2024dbl,
    author = "Geng, Hao",
    title = "{Quantum Rods and Clock in a Gravitational Universe}",
    eprint = "2412.03636",
    archivePrefix = "arXiv",
    primaryClass = "hep-th",
    month = "12",
    year = "2024"
}

@article{Geng:2024jmm,
    author = "Geng, Hao and Jiang, Yikun",
    title = "{Microscopic origin of the entropy of single-sided black holes}",
    eprint = "2409.12219",
    archivePrefix = "arXiv",
    primaryClass = "hep-th",
    doi = "10.1007/JHEP04(2025)133",
    journal = "JHEP",
    volume = "04",
    pages = "133",
    year = "2025"
}

@article{Geng:2025gqu,
    author = "Geng, Hao and Jafferis, Daniel and Shrivastava, Pushkal and Tata, Neeraj",
    title = "{The Fate of Information Localizability and Holography in Quantum Gravity}",
    eprint = "2512.18912",
    archivePrefix = "arXiv",
    primaryClass = "hep-th",
    month = "12",
    year = "2025"
}

@article{Geng:2025gns,
    author = "Geng, Hao and Huertas, Jes{\'u}s and Karch, Andreas and Randall, Lisa and Thomas, Dawson",
    title = "{Wet Hair: Global Symmetries in Entanglement Islands}",
    eprint = "2512.11025",
    archivePrefix = "arXiv",
    primaryClass = "hep-th",
    month = "12",
    year = "2025"
}

@article{Geng:2025byh,
    author = "Geng, Hao",
    title = "{Making the Case for Massive Islands}",
    eprint = "2509.22775",
    archivePrefix = "arXiv",
    primaryClass = "hep-th",
    month = "9",
    year = "2025"
}

@article{Geng:2026asi,
    author = "Geng, Hao and Karch, Andreas and Perez-Pardavila, Carlos and Raju, Suvrat and Randall, Lisa and Riojas, Marcos",
    title = "{Seeing Page Curves and Islands with Blinders On}",
    eprint = "2602.06543",
    archivePrefix = "arXiv",
    primaryClass = "hep-th",
    month = "2",
    year = "2026"
}

@article{Geng:2026wok,
    author = "Geng, Hao and Merz, Moritz and Randall, Lisa",
    title = "{Covariant Locally Localized Gravity and vDVZ Continuity}",
    eprint = "2604.28194",
    archivePrefix = "arXiv",
    primaryClass = "hep-th",
    month = "4",
    year = "2026"
}

@article{Anastasi:2025puv,
    author = "Anastasi, Edoardo and Angius, Roberta and Huertas, Jes{\'u}s and Uranga, Angel M. and Wang, Chuying",
    title = "{Relative quantum gravity: localized gravity and the swampland}",
    eprint = "2501.03310",
    archivePrefix = "arXiv",
    primaryClass = "hep-th",
    doi = "10.1007/JHEP08(2025)107",
    journal = "JHEP",
    volume = "08",
    pages = "107",
    year = "2025"
}

@article{AliAhmad:2024wja,
    author = "Ali Ahmad, Shadi and Chemissany, Wissam and Klinger, Marc S. and Leigh, Robert G.",
    title = "{Quantum reference frames from top-down crossed products}",
    eprint = "2405.13884",
    archivePrefix = "arXiv",
    primaryClass = "hep-th",
    doi = "10.1103/PhysRevD.110.065003",
    journal = "Phys. Rev. D",
    volume = "110",
    number = "6",
    pages = "065003",
    year = "2024"
}

@article{AliAhmad:2024eun,
    author = "Ali Ahmad, Shadi and Klinger, Marc S. and Lin, Simon",
    title = "{Semifinite von Neumann algebras in gauge theory and gravity}",
    eprint = "2407.01695",
    archivePrefix = "arXiv",
    primaryClass = "hep-th",
    month = "7",
    year = "2024"
}

@article{AliAhmad:2024saq,
    author = "Ali Ahmad, Shadi and Klinger, Marc S.",
    title = "{Emergent Geometry from Quantum Probability}",
    eprint = "2411.07288",
    archivePrefix = "arXiv",
    primaryClass = "hep-th",
    month = "11",
    year = "2024"
}

@article{Emparan:1999gf,
    author = "Emparan, Roberto",
    title = "{AdS / CFT duals of topological black holes and the entropy of zero energy states}",
    eprint = "hep-th/9906040",
    archivePrefix = "arXiv",
    reportNumber = "DTP-99-37, EHU-FT-9909",
    doi = "10.1088/1126-6708/1999/06/036",
    journal = "JHEP",
    volume = "06",
    pages = "036",
    year = "1999"
}

@article{Maldacena:2004rf,
    author = "Maldacena, Juan Martin and Maoz, Liat",
    title = "{Wormholes in AdS}",
    eprint = "hep-th/0401024",
    archivePrefix = "arXiv",
    reportNumber = "ITFA-2003-57",
    doi = "10.1088/1126-6708/2004/02/053",
    journal = "JHEP",
    volume = "02",
    pages = "053",
    year = "2004"
}

@article{Planck:2018vyg,
    author = "Aghanim, N. and others",
    collaboration = "Planck",
    title = "{Planck 2018 results. VI. Cosmological parameters}",
    eprint = "1807.06209",
    archivePrefix = "arXiv",
    primaryClass = "astro-ph.CO",
    doi = "10.1051/0004-6361/201833910",
    journal = "Astron. Astrophys.",
    volume = "641",
    pages = "A6",
    year = "2020",
    note = "[Erratum: Astron.Astrophys. 652, C4 (2021)]"
}

@article{Chakraborty:2023yed,
    author = "Chakraborty, Tuneer and Chakravarty, Joydeep and Godet, Victor and Paul, Priyadarshi and Raju, Suvrat",
    title = "{The Hilbert space of de Sitter quantum gravity}",
    eprint = "2303.16315",
    archivePrefix = "arXiv",
    primaryClass = "hep-th",
    doi = "10.1007/JHEP01(2024)132",
    journal = "JHEP",
    volume = "01",
    pages = "132",
    year = "2024"
}

@article{Chakraborty:2025izq,
    author = "Chakraborty, Tuneer and H, Ashik and Raju, Suvrat",
    title = "{Cosmological correlators in gravitationally-constrained de Sitter states}",
    eprint = "2507.15926",
    archivePrefix = "arXiv",
    primaryClass = "hep-th",
    doi = "10.1007/JHEP01(2026)004",
    journal = "JHEP",
    volume = "01",
    pages = "004",
    year = "2026"
}

@article{McNamara:2020uza,
    author = "McNamara, Jacob and Vafa, Cumrun",
    title = "{Baby Universes, Holography, and the Swampland}",
    eprint = "2004.06738",
    archivePrefix = "arXiv",
    primaryClass = "hep-th",
    month = "4",
    year = "2020"
}

@article{McNamara:2019rup,
    author = "McNamara, Jacob and Vafa, Cumrun",
    title = "{Cobordism Classes and the Swampland}",
    eprint = "1909.10355",
    archivePrefix = "arXiv",
    primaryClass = "hep-th",
    month = "9",
    year = "2019"
}

@article{Harlow:2025pvj,
    author = "Harlow, Daniel and Usatyuk, Mykhaylo and Zhao, Ying",
    title = "{Quantum mechanics and observers for gravity in a closed universe}",
    eprint = "2501.02359",
    archivePrefix = "arXiv",
    primaryClass = "hep-th",
    reportNumber = "MIT-CTP/5824",
    doi = "10.1007/JHEP02(2026)108",
    journal = "JHEP",
    volume = "02",
    pages = "108",
    year = "2026"
}

@article{Geng:2025rov,
    author = "Geng, Hao",
    title = "{The Mechanism behind the Information Encoding for Islands}",
    eprint = "2502.08703",
    archivePrefix = "arXiv",
    primaryClass = "hep-th",
    month = "2",
    year = "2025"
}

@article{Geng:2025bcb,
    author = "Geng, Hao and Jiang, Yikun and Xu, Jiuci",
    title = "{Algebras, Entanglement Islands, and Observers}",
    eprint = "2506.12127",
    archivePrefix = "arXiv",
    primaryClass = "hep-th",
    month = "6",
    year = "2025"
}

@article{Abdalla:2025gzn,
    author = "Abdalla, Ahmed I. and Antonini, Stefano and Iliesiu, Luca V. and Levine, Adam",
    title = "{The gravitational path integral from an observer{\textquoteright}s point of view}",
    eprint = "2501.02632",
    archivePrefix = "arXiv",
    primaryClass = "hep-th",
    doi = "10.1007/JHEP05(2025)059",
    journal = "JHEP",
    volume = "05",
    pages = "059",
    year = "2025"
}

@article{Antonini:2025sur,
    author = "Antonini, Stefano and Chen, Chang-Han and Maxfield, Henry and Penington, Geoff",
    title = "{An apologia for islands}",
    eprint = "2506.04311",
    archivePrefix = "arXiv",
    primaryClass = "hep-th",
    month = "6",
    year = "2025"
}

@article{Geng:2025efs,
    author = "Geng, Hao and Hung, Ling-Yan and Jiang, Yikun",
    title = "{It from ETH: Multi-interval Entanglement and Replica Wormholes from Large-$c$ BCFT Ensemble}",
    eprint = "2505.20385",
    archivePrefix = "arXiv",
    primaryClass = "hep-th",
    month = "5",
    year = "2025"
}

@article{Antonini:2024bbm,
    author = "Antonini, Stefano and Bariuan, Luis Gabriel C.",
    title = "{Magnetic braneworlds: cosmology and wormholes}",
    eprint = "2405.18465",
    archivePrefix = "arXiv",
    primaryClass = "hep-th",
    doi = "10.1007/JHEP09(2024)070",
    journal = "JHEP",
    volume = "09",
    pages = "070",
    year = "2024"
}

@article{Bao:2025plr,
    author = "Bao, Ning and Geng, Hao and Jiang, Yikun",
    title = "{Ryu-Takayanagi formula for multi-boundary black holes from 2D large-c CFT ensemble}",
    eprint = "2504.12388",
    archivePrefix = "arXiv",
    primaryClass = "hep-th",
    doi = "10.1007/JHEP10(2025)042",
    journal = "JHEP",
    volume = "10",
    pages = "042",
    year = "2025"
}

@article{Chua:2023ios,
    author = "Chua, Wan Zhen and Jiang, Yikun",
    title = "{Hartle-Hawking state and its factorization in 3d gravity}",
    eprint = "2309.05126",
    archivePrefix = "arXiv",
    primaryClass = "hep-th",
    doi = "10.1007/JHEP03(2024)135",
    journal = "JHEP",
    volume = "03",
    pages = "135",
    year = "2024"
}

@article{Chen:2024unp,
    author = "Chen, Lin and Hung, Ling-Yan and Jiang, Yikun and Lao, Bing-Xin",
    title = "{Deriving the non-perturbative gravitational dual of quantum Liouville theory from BCFT operator algebra}",
    eprint = "2403.03179",
    archivePrefix = "arXiv",
    primaryClass = "hep-th",
    doi = "10.21468/SciPostPhys.19.6.163",
    journal = "SciPost Phys.",
    volume = "19",
    number = "6",
    pages = "163",
    year = "2025"
}

@article{Hung:2024gma,
    author = "Hung, Ling-Yan and Jiang, Yikun",
    title = "{Building up quantum spacetimes with BCFT Legos}",
    eprint = "2404.00877",
    archivePrefix = "arXiv",
    primaryClass = "hep-th",
    month = "3",
    year = "2024"
}

@article{Bao:2024ixc,
    author = "Bao, Ning and Hung, Ling-Yan and Jiang, Yikun and Liu, Zhihan",
    title = "{QG from SymQRG: AdS$_3$/CFT$_2$ Correspondence as Topological Symmetry-Preserving Quantum RG Flow}",
    eprint = "2412.12045",
    archivePrefix = "arXiv",
    primaryClass = "hep-th",
    month = "12",
    year = "2024"
}

@article{Liu:2026fnd,
    author = "Liu, Hong",
    title = "{Ramp, Plateau, and Wormholes without Averaging, and Hyper-non-perturbative Structures in Gravity}",
    eprint = "2608.02743",
    archivePrefix = "arXiv",
    primaryClass = "hep-th",
    reportNumber = "MIT-CTP/6074",
    month = "8",
    year = "2026"
}

@article{Liu:2025ikq,
    author = "Liu, Hong",
    title = "{''Filtering'' CFTs at large N: Euclidean Wormholes, Closed Universes, and Black Hole Interiors}",
    eprint = "2512.13807",
    archivePrefix = "arXiv",
    primaryClass = "hep-th",
    reportNumber = "MIT-CTP/5970",
    month = "12",
    year = "2025"
}

@article{Liu:2025cml,
    author = "Liu, Hong",
    title = "{Towards a holographic description of closed universes}",
    eprint = "2509.14327",
    archivePrefix = "arXiv",
    primaryClass = "hep-th",
    reportNumber = "MIT-CTP/5923",
    month = "9",
    year = "2025"
}

@article{Cirafici:2026iek,
    author = "Cirafici, Michele",
    title = "{Observers, local measurements, and topology}",
    eprint = "2607.15007",
    archivePrefix = "arXiv",
    primaryClass = "hep-th",
    month = "7",
    year = "2026"
}

@article{Das:2026npy,
    author = "Das, Suchetan",
    title = "{Modular quantization and black holes}",
    eprint = "2606.11984",
    archivePrefix = "arXiv",
    primaryClass = "hep-th",
    month = "6",
    year = "2026"
}

@article{Cheung:2026euf,
    author = "Cheung, Clifford and Sivaramakrishnan, Allic and Wilson-Gerow, Jordan and Zhou, Lihang",
    title = "{On Perturbatively Dressed Observables}",
    eprint = "2605.26077",
    archivePrefix = "arXiv",
    primaryClass = "hep-th",
    month = "5",
    year = "2026"
}

@article{Seo:2026kxs,
    author = "Seo, Min-Seok",
    title = "{Implication of dressed form of relational observable on von Neumann algebra}",
    eprint = "2603.25990",
    archivePrefix = "arXiv",
    primaryClass = "hep-th",
    month = "3",
    year = "2026"
}

@article{Kaushal:2025wbn,
    author = "Kaushal, Anurag and Prabhakar, Naveen S. and Wadia, Spenta R.",
    title = "{Gauge invariant Hamiltonian evolution across the black hole horizon in asymptotically AdS spacetimes}",
    eprint = "2510.21920",
    archivePrefix = "arXiv",
    primaryClass = "hep-th",
    doi = "10.1103/gz2s-mqk5",
    journal = "Phys. Rev. D",
    volume = "113",
    number = "10",
    pages = "106019",
    year = "2026"
}

@article{Chakraborty:2025baw,
    author = "Chakraborty, Avik and Ghosh, Jewel Kumar and Molina, Mart{\'\i}n and Mukhopadhyay, Ayan",
    title = "{Junctions, strings, clocks and gravitational memory in three dimensional dS space}",
    eprint = "2510.17953",
    archivePrefix = "arXiv",
    primaryClass = "hep-th",
    month = "10",
    year = "2025"
}

@article{Ciambelli:2025ztm,
    author = "Ciambelli, Luca and Kowalski-Glikman, Jerzy and Varrin, Ludovic",
    title = "{Entanglement entropy of quantum corners}",
    eprint = "2507.16800",
    archivePrefix = "arXiv",
    primaryClass = "hep-th",
    doi = "10.1007/JHEP02(2026)188",
    journal = "JHEP",
    volume = "02",
    pages = "188",
    year = "2026"
}

@article{Aalsma:2025bcg,
    author = "Aalsma, Lars and Bak, Sang-Eon",
    title = "{Modular fluctuations in cosmology}",
    eprint = "2503.04886",
    archivePrefix = "arXiv",
    primaryClass = "hep-th",
    doi = "10.1103/cpyr-w532",
    journal = "Phys. Rev. D",
    volume = "112",
    number = "2",
    pages = "026017",
    year = "2025"
}

@article{Guijosa:2025mwh,
    author = "Guijosa, Alberto",
    title = "{On the underlying nonrelativistic nature of relativistic holography}",
    eprint = "2502.03031",
    archivePrefix = "arXiv",
    primaryClass = "hep-th",
    doi = "10.1007/JHEP08(2025)139",
    journal = "JHEP",
    volume = "08",
    pages = "139",
    year = "2025"
}

@article{AliAhmad:2025ukh,
    author = "Ali Ahmad, Shadi and Jefferson, Ro",
    title = "{Algebraic perturbation theory: traversable wormholes and generalized entropy beyond subleading order}",
    eprint = "2501.01487",
    archivePrefix = "arXiv",
    primaryClass = "hep-th",
    month = "1",
    year = "2025"
}

@article{Jensen:2024dnl,
    author = "Jensen, Kristan and Raju, Suvrat and Speranza, Antony J.",
    title = "{Holographic observers for time-band algebras}",
    eprint = "2412.21185",
    archivePrefix = "arXiv",
    primaryClass = "hep-th",
    doi = "10.1007/JHEP06(2025)242",
    journal = "JHEP",
    volume = "06",
    pages = "242",
    year = "2025"
}

@article{Piazza:2024lpt,
    author = "Piazza, Federico and Vareilles, Sim{\'e}on",
    title = "{Cosmological perturbations meet Wheeler DeWitt}",
    eprint = "2412.19782",
    archivePrefix = "arXiv",
    primaryClass = "hep-th",
    doi = "10.1088/1475-7516/2025/07/061",
    journal = "JCAP",
    volume = "07",
    pages = "061",
    year = "2025"
}

@article{Cirafici:2024itu,
    author = "Cirafici, Michele",
    title = "{Gravitational Algebras and Applications to Nonequilibrium Physics}",
    eprint = "2412.17674",
    archivePrefix = "arXiv",
    primaryClass = "hep-th",
    doi = "10.3390/universe11010024",
    journal = "Universe",
    volume = "11",
    number = "1",
    pages = "24",
    year = "2025"
}

@article{Grassi:2024vkb,
    author = "Grassi, Pietro Antonio and Porrati, Massimo",
    title = "{Local operator algebras of charged states in gauge theory and gravity}",
    eprint = "2411.08865",
    archivePrefix = "arXiv",
    primaryClass = "hep-th",
    doi = "10.1007/JHEP03(2025)175",
    journal = "JHEP",
    volume = "03",
    pages = "175",
    year = "2025"
}

@article{McNamara:2026isz,
    author = "McNamara, Jacob and Wang, Zhencheng",
    title = "{Wormholes as red herrings: reflection positivity and the reconstruction of unitary quantum field theories}",
    eprint = "2607.01322",
    archivePrefix = "arXiv",
    primaryClass = "hep-th",
    month = "7",
    year = "2026"
}

@article{Blommaert:2025bgd,
    author = "Blommaert, Andreas and Kudler-Flam, Jonah and Urbach, Erez Y.",
    title = "{Absolute entropy and the observer{\textquoteright}s no-boundary state}",
    eprint = "2505.14771",
    archivePrefix = "arXiv",
    primaryClass = "hep-th",
    doi = "10.1007/JHEP11(2025)113",
    journal = "JHEP",
    volume = "11",
    pages = "113",
    year = "2025"
}

@article{Blommaert:2026lvp,
    author = "Blommaert, Andreas and Chen, Chang-Han",
    title = "{Time in gravitational subregions and in closed universes}",
    eprint = "2602.22153",
    archivePrefix = "arXiv",
    primaryClass = "hep-th",
    month = "2",
    year = "2026"
}

@article{Blommaert:2026ofx,
    author = "Blommaert, Andreas and Tietto, Damiano and Verlinde, Herman",
    title = "{An observer's quantization of 3d de Sitter}",
    eprint = "2606.26241",
    archivePrefix = "arXiv",
    primaryClass = "hep-th",
    month = "6",
    year = "2026"
}

@article{Klinger:2023tgi,
    author = "Klinger, Marc S. and Leigh, Robert G.",
    title = "{Crossed products, extended phase spaces and the resolution of entanglement singularities}",
    eprint = "2306.09314",
    archivePrefix = "arXiv",
    primaryClass = "hep-th",
    doi = "10.1016/j.nuclphysb.2024.116453",
    journal = "Nucl. Phys. B",
    volume = "999",
    pages = "116453",
    year = "2024"
}

@article{Klinger:2023auu,
    author = "Klinger, Marc S. and Leigh, Robert G.",
    title = "{Crossed products, conditional expectations and constraint quantization}",
    eprint = "2312.16678",
    archivePrefix = "arXiv",
    primaryClass = "hep-th",
    doi = "10.1016/j.nuclphysb.2024.116622",
    journal = "Nucl. Phys. B",
    volume = "1006",
    pages = "116622",
    year = "2024"
}

@article{Klinger:2025hjp,
    author = "Klinger, Marc S. and Leigh, Robert G.",
    title = "{The problem of time and its quantum resolution}",
    eprint = "2504.00152",
    archivePrefix = "arXiv",
    primaryClass = "hep-th",
    doi = "10.1142/S0218271825440225",
    journal = "Int. J. Mod. Phys. D",
    volume = "34",
    number = "16",
    pages = "2544022",
    year = "2025"
}

@article{Buchel:2004rr,
    author = "Buchel, Alex",
    title = "{Gauge theories on hyperbolic spaces and dual wormhole instabilities}",
    eprint = "hep-th/0402174",
    archivePrefix = "arXiv",
    doi = "10.1103/PhysRevD.70.066004",
    journal = "Phys. Rev. D",
    volume = "70",
    pages = "066004",
    year = "2004"
}

@article{Baumann:2026lnt,
    author = "Baumann, Daniel and Hinterbichler, Kurt and Jones, Callum R. T. and Meurrens, Nathan",
    title = "{Bootstrapping Weakly Broken Gauge Theories in (A)dS}",
    eprint = "2609.38308",
    archivePrefix = "arXiv",
    primaryClass = "hep-th",
    month = "9",
    year = "2026"
}

@misc{dai2026babyuniverselargefamily,
      title={A baby universe from a large family: booklet cosmology states and quantum error correction}, 
      author={Jingshu Dai and Binye Dong and Cheng Peng},
      year={2026},
      eprint={2610.02168},
      archivePrefix={arXiv},
      primaryClass={hep-th},
      url={https://arxiv.org/abs/2610.02168}, 
}

@misc{giddings2026leadinggravitationaldressingoperators,
      title={Leading gravitational dressing of operators and states in de Sitter space}, 
      author={Steven B. Giddings and Zi-Yue Wang},
      year={2026},
      eprint={2610.02209},
      archivePrefix={arXiv},
      primaryClass={hep-th},
      url={https://arxiv.org/abs/2610.02209}, 
}

@article{Callebaut:2026hso,
    author = "Callebaut, Nele and Selle, Matteo",
    title = "{Brane effective actions and their island rule from $T\overline T$ flows}",
    eprint = "2608.30533",
    archivePrefix = "arXiv",
    primaryClass = "hep-th",
    month = "8",
    year = "2026"
}

\end{document}